\documentclass[%
 aip,
 amsmath,amssymb,
 reprint,%
]{revtex4-1}
 \usepackage[pdftex]{graphicx}
\usepackage{hyperref}
\usepackage{xcolor}
\usepackage{bm}
\usepackage{amsmath}
\usepackage{dcolumn}

\makeatletter
\def\@email#1#2{%
 \endgroup
 \patchcmd{\titleblock@produce}
  {\frontmatter@RRAPformat}
  {\frontmatter@RRAPformat{\produce@RRAP{*#1\href{mailto:#2}{#2}}}\frontmatter@RRAPformat}
  {}{}
}%
\makeatother

\begin{document}

\preprint{AIP/123-QED}

\title{Estimation of Exciton Binding Energy and lifetime for Mono-layer Transition Metal Dichalcogenides}
\author{Rohit Ramesh Nimje} 
\author{ Swati G}
\author{ Ashutosh Mahajan}
\affiliation{Centre for Nanotechnology Research, Vellore Institute of Technology, Vellore-632 014, India}
 
\email{ashutosh.mahajan@vit.ac.in}

\date{\today}

\begin{abstract}

In this work, we present a mathematical model for the  Wannier-Mott exciton in monolayers of transition metal dichalcogenides such as $WS_2$, $WSe_2$, $MoS_2$, $MoSe_2$ that estimates the radiation lifetime in the effective mass approximation. We calculate exciton energy, and binding energy by solving the Schrodinger wave equation with open boundary conditions to obtain quasi-bound states in the confined direction in the monolayer and decay rates by the Fermi-Golden rule. The proposed model uses only the physical parameters such as band offsets, effective mass, and dielectric constants for the monolayers of $WS_2$, $WSe_2$, $MoS_2$, and $MoSe_2$. The model is validated against III-V material quantum well heterostructure, and the estimated effective lifetime considering the thermalization of the exciton has been compared with photoluminescence decay for the TMD heterostructure. Our calculated values show good agreement with the time-resolved photoluminescence spectroscopy measurements and DFT estimations.
\end{abstract}
\maketitle
\section{Introduction}

Monolayer transition metal dichalcogenides (TMDs) possess immense potential for optoelectronic device applications \cite{mueller2018exciton,ahmed2017two,liu2015chemical,lee2014atomically,baugheroptoelectronics,yin2012single}, owing to quantum confinement and reduced dielectric screening. The direct excitons in monolayer TMDs, such as $WS_2$, $WSe_2$, $MoS_2$, and $MoSe_2$, exhibit longer lifetimes compared to their bulk counterparts, attracting significant research interest in recent years. These materials display remarkable optical properties resulting from strong light$–$matter interactions and their direct bandgaps. 
Due to their significantly lower dielectric constants compared to bulk forms, monolayer TMDs exhibit high exciton binding energies—typically in the range of 100$–$500 meV \cite{zhang2015excited}$—$which confer exceptional exciton stability and enable the optical characteristics to persist even at room temperature. 

Direct excitons in TMD monolayers decay slowly, with photoluminescence (PL) lifetimes reported in the range of 220$–$420 picoseconds \cite{mohamed2017long}. Excitons play a pivotal role in various advanced applications, including valleytronics \cite{gong2013magnetoelectric,jiang2018microsecond}, quantum computing \cite{liu20192d}, biosensing \cite{gan2017two}, ultrasensitive visible \cite{lopez2013ultrasensitive} and infrared \cite{lukman2020high} photodetectors, light$-$emitting diodes \cite{ross2014electrically,pospischil2014solar}, and single$-$photon emitters \cite{koperski2015single,he2015single} and exciton lifetime is a crucial parameter in most of these applications. The monolayer $WS_2$ quantum structures have been explored to measure exciton and trion energies in photonics devices \cite{sun2023enhanced}, while monolayer $MoS_2$ and $MoS_2$ has been utilized in spintronic \cite{eknapakul2014electronic}, WSe2  \cite{} for and MoSe2 for \cite{} and other information technology platforms \cite{qiu2024two}.

Understanding the unique properties of excitons in TMDs is essential both for advancing fundamental physics and for enabling new technologies. Various computational approaches have been developed to estimate excitonic states and binding energies in monolayer TMDs, including density functional theory (DFT) \cite{deng2018stability,palummo2015exciton}, atomistic tight-binding models \cite{wang2016radiative,berghauser2014analytical}, and effective mass approximations (EMA) \cite{berghauser2014analytical,cheiwchanchamnangij2012quasiparticle}. Several theoretical investigations have been reported on exciton dynamics in 2D TMD heterostructures. Martins {\it et al.} employed a variationally optimized two-dimensional modified Laguerre basis set combined with exact diagonalization \cite{wu2019exciton}, based on Wannier-Mott exciton theory \cite{martins2020colloquium}. 
The stochastic variational method (SVM) has proven effective for exciton binding energy calculations due to its adaptable wavefunction representation, which accommodates inter$-$particle interactions \cite{zhang2015excited}. First$-$principles DFT calculations are typically employed to derive electronic band structures, followed by GW approximations—accounting for many-body electron–hole interactions—to correct the bandgap \cite{palummo2015exciton,kronik2016excited}. 

Experimental studies on h$-$BN$-$stacked $WS_2$ monolayers and multilayer heterostructures have reported exciton energy measurements using time-resolved PL spectroscopy in resonant tunneling diode (RTD) structures \cite{lee2020measurement}. These findings demonstrate a reduction in the bandgap with increasing $WS_2$ layer thickness, even up to four layers. 
The presence of a single h-BN layer also suppresses interlayer excitonic interactions in multilayer $WS_2$ structures under an applied electric field \cite{lee2020measurement}. Time$-$resolved photoluminescence (TRPL) studies have shown exciton decay times of approximately 806 ps for monolayers, 401 ps for bilayers, and 332 ps for trilayers in $WS_2$, highlighting the influence of dielectric screening, temperature, and thickness on exciton dynamics \cite{yuan2015exciton}.

Exciton energies have been measured experimentally in $WS_2$ \cite{zhu2015exciton,chernikov2014exciton,chernikov2015electrical}, and both exciton and trion energies have been reported for $MoS_2$ and $MoSe_2$ in Ref. \cite{lin2014dielectric,wang2022disorder,lee2020measurement}. 
Exciton lifetimes in monolayer TMDs range from 2.62 to 5.7 picoseconds at low temperatures \cite{mohamed2017long}, and from 0.22 to 0.42 nanoseconds at room temperature \cite{palummo2015exciton}. The significant increase in exciton lifetime at room temperature is attributed to thermalization, which broadens the distribution of in$-$plane momenta of exciton energy states \cite{wang2016radiative}.

A comprehensive model to estimate lifetime variation with respect to material properties, geometry, and temperature is highly desirable. The theoretical model employed in this study for the h$-$$BN$$/$$WS_2$$/$h$-$BN heterostructure involves solving the Schrödinger equation for a single quantum well using the EMA as reported in our previous works \cite{pscripta2023analytical,mahajan2021analytical}. This work facilitates the prediction of exciton energy, binding energy, and radiative lifetime for monolayer TMDs, which are essential parameters in the design and engineering of materials for optoelectronic applications. The effective mass method provides a straightforward yet powerful approach to compute exciton energy states in low-dimensional quantum systems.


\section{Mathematical Model}
\begin{figure}[ht]
\includegraphics[width=8.5cm]{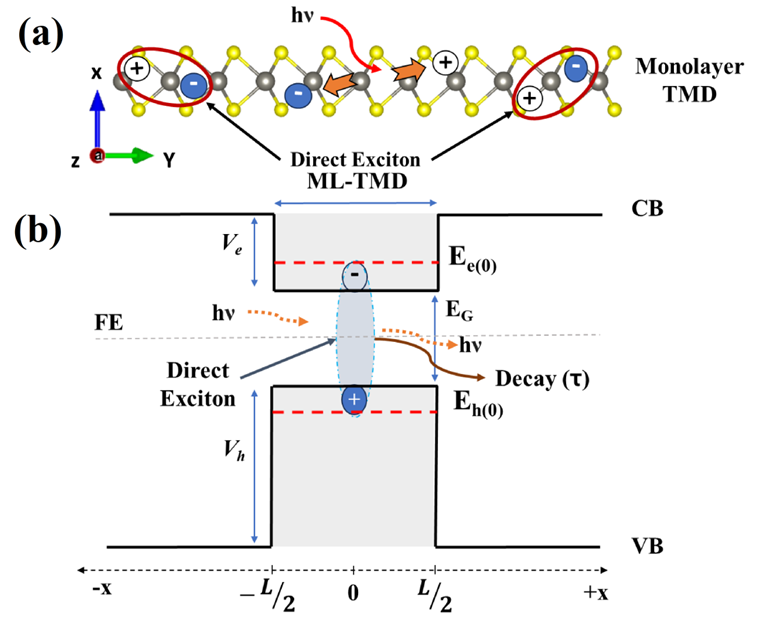}
\par\caption{(a) Schematic representation of Direct exciton in monolayer TMDs; (b) Band diagram indicating exciton formation and exciton decay.} 
\label{fig1}
\end{figure}

The excitonic Hamiltonian can be written for the system depicted in Fig. \ref{fig1} as
\begin{equation}
H=H_e+H_h+E_g+T_X-\frac{q^2}{\epsilon |r_e-r_h|}
\label{eq1a}
\end{equation}
where, $H_{e/h}$ is the effective mass Hamiltonian for the electron/hole in conduction/valance band in confined direction, $E_g$ is the bandgap of TMD material, $T_X$ is the kinetic energy term for the in-plane relative motion, and the last term is the coloumb interaction potential.
The one-dimensional time-independent Schr\"odinger equation along the device direction can be written for electron and hole in the presence of electric field $\xi$ as
\begin{align}
 \frac{\partial^{2} \psi_{e,h}}{\partial x_{e,h}^{2}}-\frac{2 m_{e,h}}{\hbar^2}  \left(\xi x_{e,h}-{E_{e,h}}\right) \psi_{e,h}=0, \hspace{0.5cm} x_{e.h}\le \left| \frac{L}{2}\right|  \nonumber\\
\frac{\partial^{2} \psi_{e,h}}{\partial x_{e,h}^{2}}-\frac{2 m_{e,h}}{\hbar^2}\left(\xi x_{e,h}+{V_{e,h}-E_{e,h}}\right) \psi_{e,h}=0, \hspace{0.5cm}  x_{e,h} > \frac{L}{2} , \nonumber\\
\label{eq2a}
\end{align}
The excitonic wavefunction for the heterostructure in the cylindrical coordinate system can be written as \cite{belov2019energy}
\begin{eqnarray}
\Psi(x_e,x_h,\rho) = {\psi}_e(x_e) {\psi}_h(x_h)\frac{\psi(\rho)}{\rho}e^{il\phi}
\label{eq3a}
\end{eqnarray}
where, $\psi_e$ and $\psi_h$ are the electron and hole wavefunctions for the Hamiltonian of the 1D quantum well as shown in Fig.  \ref{fig1} (b), $x_e$ and $x_h$ are the coordinates along the growth direction, and $m_e$ and $m_h$ are the effective masses. $\Psi$ is part of the wavefunction in the other two directions in the cylindrical coordinate system. Since mainly the 1s state gives prominent PL emission, we take $l=0$.

The boundary conditions of the above equation (\ref{eq2a}) are open, therefore, the eigen energies are complex 
\begin{eqnarray}
E_{e,h} = E_{e,h}^0 - i\Gamma_{e,h}^0 \nonumber
\end{eqnarray}

The ground state of the electron and hole within the conduction band and valence band has been considered to form a direct exciton, as shown in Fig.  \ref{fig1}(b).
The Wannier equation for the cylindrical part of the wave function can be written as
\begin{align}
\left( \frac{{\partial^2}}{{\partial \rho^2}}+ \frac{1}{\rho} \frac{\partial}{\partial \rho} \right)\psi(\rho) +\frac{2\mu}{\hbar^2} \left[V_{C}-( E_n^{0}-E_X) \right] \psi(\rho) = 0  
\label{eq4a}
\end{align}    
 $E_X$ denotes the exciton eigenenergy, and $E_{n}$$^0$ is the energy difference between the electron and hole states' energies in equation (\ref{eq2a}) as shown in Fig.  \ref{fig3}.
The exciton-reduced mass can be written as
\begin{equation}
    \frac{1}{\mu}= \frac{1}{m_e} + \frac{1}{m_h} 
    \label{eq5a}
\end{equation}

$E_{n}^0$ can be written as below   
\begin{eqnarray}    
E_{n}^{0}=E_{e}^{0}+E_{h}^{0}+E_g 
\label{eq6a}
\end{eqnarray}
and binding energy can be written as 
\begin{eqnarray}    
E_{b}=E_{n}^{0}-E_{X} 
\label{eq7a}
\end{eqnarray}

The Coulomb interaction between the electron and hole is given as 
\begin{eqnarray}
V_C(\rho) = -\frac{e^2}{\epsilon_b} \int_{x_{e_{1}}}^{x_{e_{2}}}\int_{x_{h_{1}}}^{x_{h_{2}}}  \frac{{\psi}_e(x_e) {\psi}_h(x_h)}{\sqrt{(x_e - x_h)^2 + \rho^2}} dx_edx_h
\label{eq8a}
\end{eqnarray}
where $\epsilon_b$ is the dielectric constant in monolayer TMDs. The $\epsilon_b$ is obtained from the first$-$principle DFT calculations at room temperature \cite{laturia2018dielectric}, assuming zero epitaxial strain as seen in table \ref{table1}. \\


\subsection{Exciton Decay Lifetime}
Radiative lifetime of direct exciton depends on the oscillator strength $f_{val}$, which is a measure of the strength of light-matter interaction.
The oscillator strength can be derived by considering the probability of electromagnetic dipole transitions occurring, where the initial state is excitonic, and the final state is the crystal ground state and a photon. The oscillator strength can be written as \cite{burstein2012confined}
\begin{equation}
f_{val}=\frac{2}{m \hbar\omega}|<\Psi_0|\epsilon\cdot p|\Psi_{ex}>|^2
\label{eq9a}
\end{equation}
where, $\Psi_0$ is the crystal ground state,$\Psi_{ex}$ is the exciton state, $\epsilon$ is the polarization vector and $p$ is pseudomomentum operator for the electron. The overlap of the electron and hole envelope functions determines the strength of the transition.

Since for the direct transition, the electron and hole momentum vectors are related as $k_e=-k_h$, the above expression can be further worked out as \cite{dimmock1967introduction}
\begin{equation}
f_{val}=\frac{2}{ m \hbar\omega}|<u_v|\epsilon\cdot p|u_c>|^2 \left| \int \Psi(\rho=0,x_e,x_h)dx_edx_h\right|^2
\label{eq10a}
\end{equation}
where, $u_c$ \& $u_v$ are periodic parts of the Bloch functions for conduction and valence bands.

 The oscillator strength can be written in terms of the transition dipole moment $d_{cv}$ between the valence and the conduction band Bloch functions as,
\begin{equation}
        f_{val} = \frac{2m_0 E_{X} d_{cv}^2}{\hbar^2} \left| \int_{x_{e_{1}}}^{x_{e_{2}}}\int_{x_{h_{1}}}^{x_{h_{2}}}\Psi(x_e,x_h,\rho=0)dx_edx_h\right|^2
    \label{eq11a}
\end{equation}

The intrinsic radiative recombination time, or the zero in-plane momentum excitons, has extremely short radiative lifetimes and can be written as 
\begin{equation}
    \tau=\frac{\hbar}{2\Gamma_0}
    \label{eq12a}
\end{equation}
where, exciton radiative linewidth can be written in terms of the oscillator strength as \cite{citrin1993radiative}

\begin{equation}
    \Gamma_0 = \frac{\pi e^2}{\sqrt{\epsilon_{b}} m_0 c} \cdot f_{val}
\label{eq13a}
\end{equation}

The intrinsic lifetime calculated above turns out to be much smaller than the experimentally observed lifetimes by time-resolved photoluminescence spectroscopy \cite{mohamed2017long}, therefore, the thermalization process of the exciton must be taken into account.

\begin{subsection}{Thermalization}

The exciton decay time at room temperature can further be calculated by incorporating a thermalization factor to the radiative lifetime as given in \cite{robert2016exciton}. This thermalization factor is due to the inelastic scattering of excitons with the acoustic phonons, which does not change the exciton population. The thermalization process is much faster than the radiative recombination. Therefore, excitons will always have a thermal distribution when the radiative recombination process happens.

The observed decay rate at a given temperature can be given as an average over the thermal distribution given by the Maxwell-Boltzmann distribution \cite{rowlinson2005maxwell}.
\begin{equation}
<\Gamma_{eff}>=\frac{\sum_s \Gamma_s e^{-\beta{E_S}}}{\sum_s e^{-\beta{E_S} }}
\label{eq14a}
\end{equation}

The dispersion relation for excitons is given as \cite{wang2016radiative} \cite{dimmock1967introduction}

\begin{eqnarray}
E_s(k)=E_s(0)+\frac{\hbar^2k^2}{2M} \nonumber    
\end{eqnarray}
where, M=$m_e$+$m_h$ and $k$ is exciton momentum. For a given in-plane wavevector, orthogonal photon polarization vector can be chosen as
$\epsilon^1=\hat{y}$ and $\epsilon^2=k_z(\hat{x}-\hat{z})$. The decay widths in the longitudinal and transverse directions can be evaluated as

\begin{equation}
\Gamma_{L}=2\Gamma_0 \frac{\int_0^{k_0} k_zdk_z\frac{\sqrt{k_0^2-k^2}}{k_z} e^{-\beta{\frac{\hbar^2k_z^2}{2M}}}}{\int_0^{k_0} k_zdk_z e^{-\beta{\frac{\hbar^2k_z^2}{2M}} }} 
    \label{eq15a}
\end{equation}

\begin{equation}
\Gamma_{T}=2\Gamma_0 \frac{\int_0^{k_0} k_z dk_z\frac{k_0}{\sqrt{k_0^2-k_z^2}} e^{-\beta{\frac{\hbar^2k_z^2}{2M}}}}{\int_0^{k_0} k_zdk_z e^{-\beta{\frac{\hbar^2k_z^2}{2M}} }} 
    \label{eq16a}
\end{equation}

 Since the four spin states of excitons are equally populated, the ensemble averaged decay rate for the heavy-hole exciton can be written as \cite{burstein2012confined}
\begin{equation}
\tau_{eff}(T)=\frac{3}{2} \frac{2Mk_B T}{\hbar^2 k_o^2} \tau
    \label{eq17a}
\end{equation}


where, the light wavevector is
\begin{eqnarray}
k_0= \frac{E_{X}\sqrt{\epsilon_b}}{\hbar c}  \nonumber
\end{eqnarray}

\end{subsection}
\section {Numerical Simulations}

The eigenvalue equation (\ref{eq2a}) for electrons and holes are solved numerically using the Quantum Transmitting Boundary Method (QTBM) \cite{lent1990quantum,shao1995eigenvalue}. This approach of solving schr\"odinger wave equation in the potential barrier with an open$-$boundary condition is dependent only on the device dimensions and 2D material parameters.  The electron and hole quasi-bound states obtained by solving the 1D finite element assembled matrices for the quantum wells can be seen in Fig. \ref{fig3}. The real part of the eigenvalues is considered for the exciton energy and binding energy estimation. 
Equation (\ref{eq4a}) is assembled in a Finite difference scheme with 300 discrete points and vanishing boundary conditions are applied at $\rho=0$ and the other end. Integration for the coulomb potential in equation \ref{eq8a} is carried out using the trapezoidal rule. Numerical codes are implemented in MATLAB.
\subsection{Model Validation}
\begin{figure}[ht]
\includegraphics[width=8.5cm]{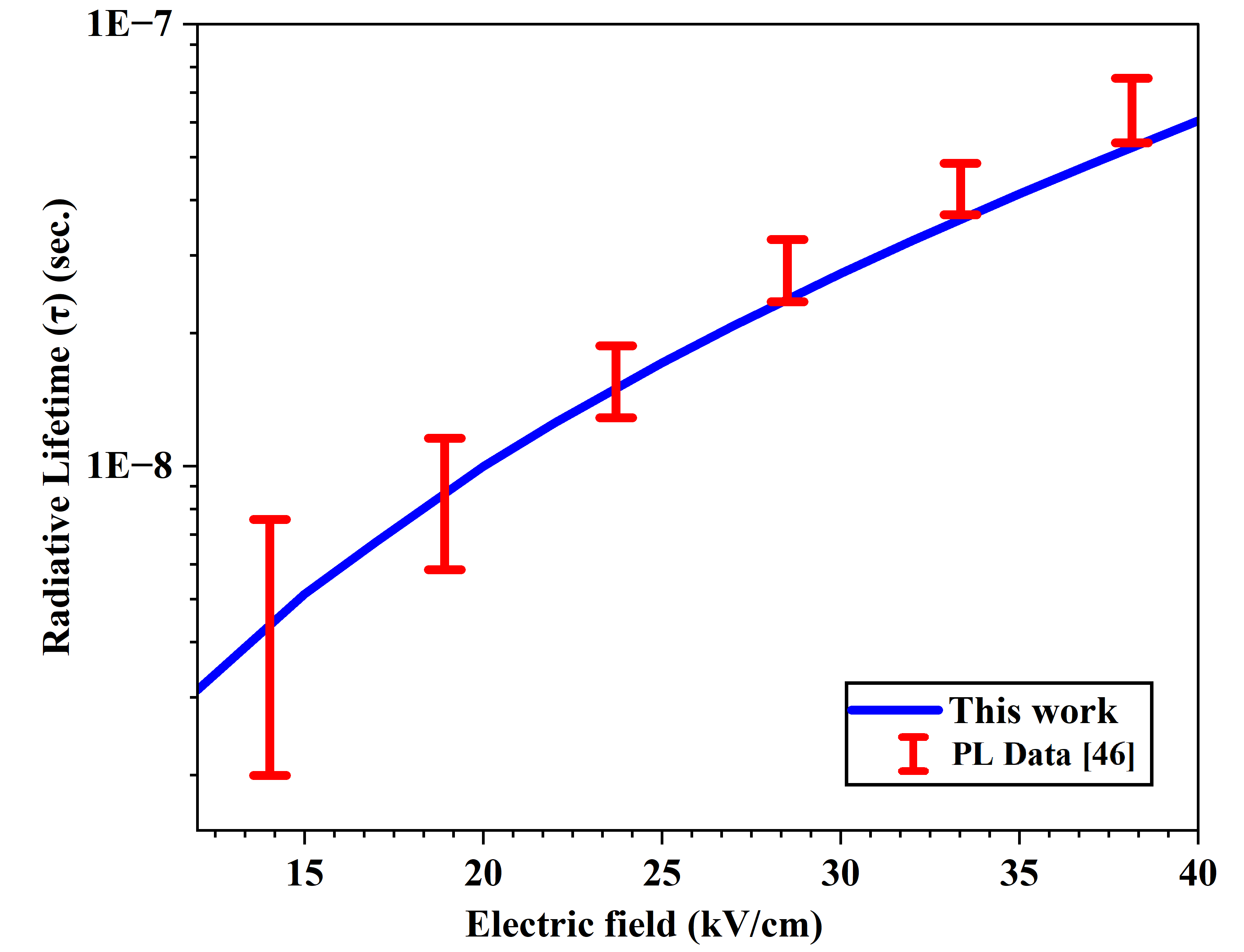}
\par\caption{Validation of the Exciton radiative decay lifetime in III$-$V double quantum well 
PL measured data \cite{butov1999photoluminescence}.}
\label{fig2}
\end{figure}

Before running the codes for the TMDs, the model validation of decay lifetime is carried out with III$-$V semiconductor heterostructure Double Quantum Well (DQW)  under an applied electric field. The calculations in \cite{sivalertporn2012direct} and experimental measurements (Ref \cite{butov1999photoluminescence}) are compared with our model calculations for validation purposes. As shown in Fig. \ref{fig2}, the calculated exciton lifetimes match quantitatively with the low-temperature photoluminescence measurements \cite{butov1999photoluminescence}, reinforcing the accuracy of our model and the numerical codes.

\subsection{Exciton quasi$-$bound state in monolayer TMDs}

We consider a single quantum well structure as seen in  Fig.  \ref{fig1} and as given in Ref. \cite{lee2020measurement}, where photoluminescence measurements of monolayer $WS_2$  are available. A single layer of TMD with a thickness as given in Table \ref{table1} is taken into account for QBS estimation.


\begin{figure}[ht]
\includegraphics[width=8.5cm]{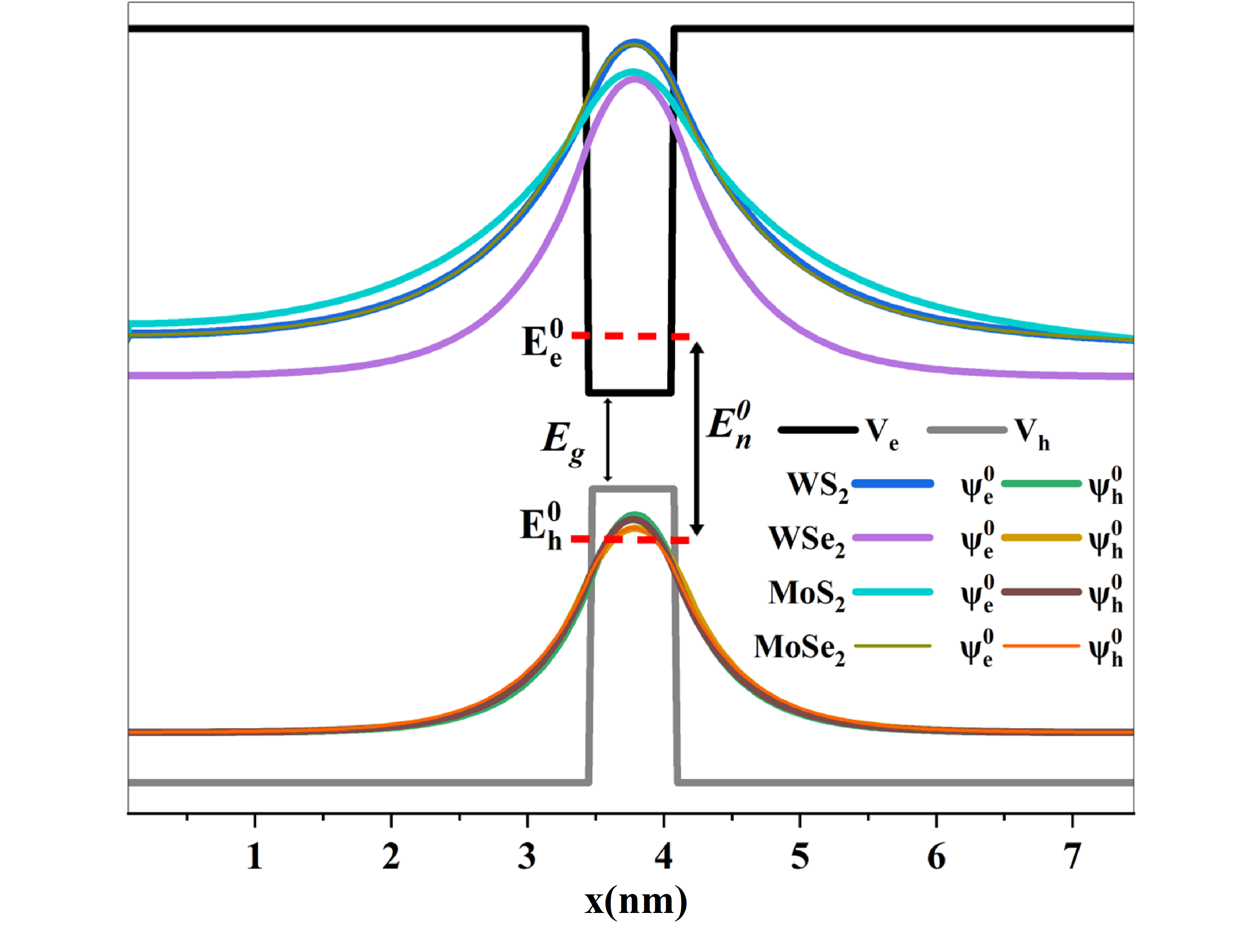}
\par\caption{Quasi$-$bound state wavefunctions for electron and hole in different monolayer TMDs.}
\label{fig3}
\end{figure}
We take vanishing boundary conditions for the excitonic wavefunction at $\rho=$0, and $\rho=$23.70 nm.
  The $\epsilon_b$ for monolayers is different than bulk; this can change the coulomb interaction shown in equation \ref{fig3} between electron and hole. The effective mass $\mu$ of the exciton strongly affects the wavefunction spread and modifies the recombination rate.

\begin{figure}[ht]
\includegraphics[width=8.5cm]{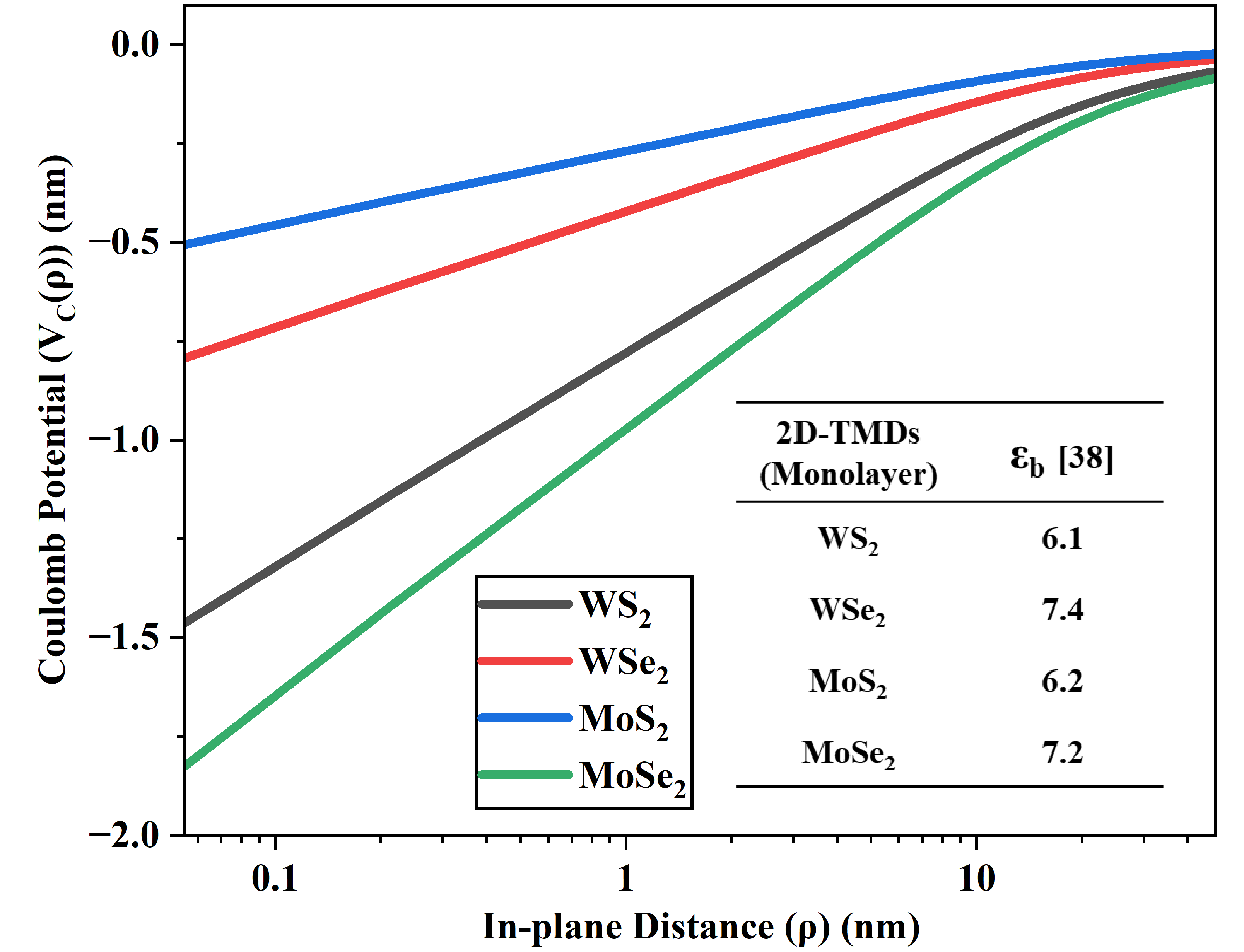}
\par\caption{Coulomb potential for different monolayer TMDs}
\label{fig4}
\end{figure}

\begin{figure}[ht]
\includegraphics[width=8.5cm]{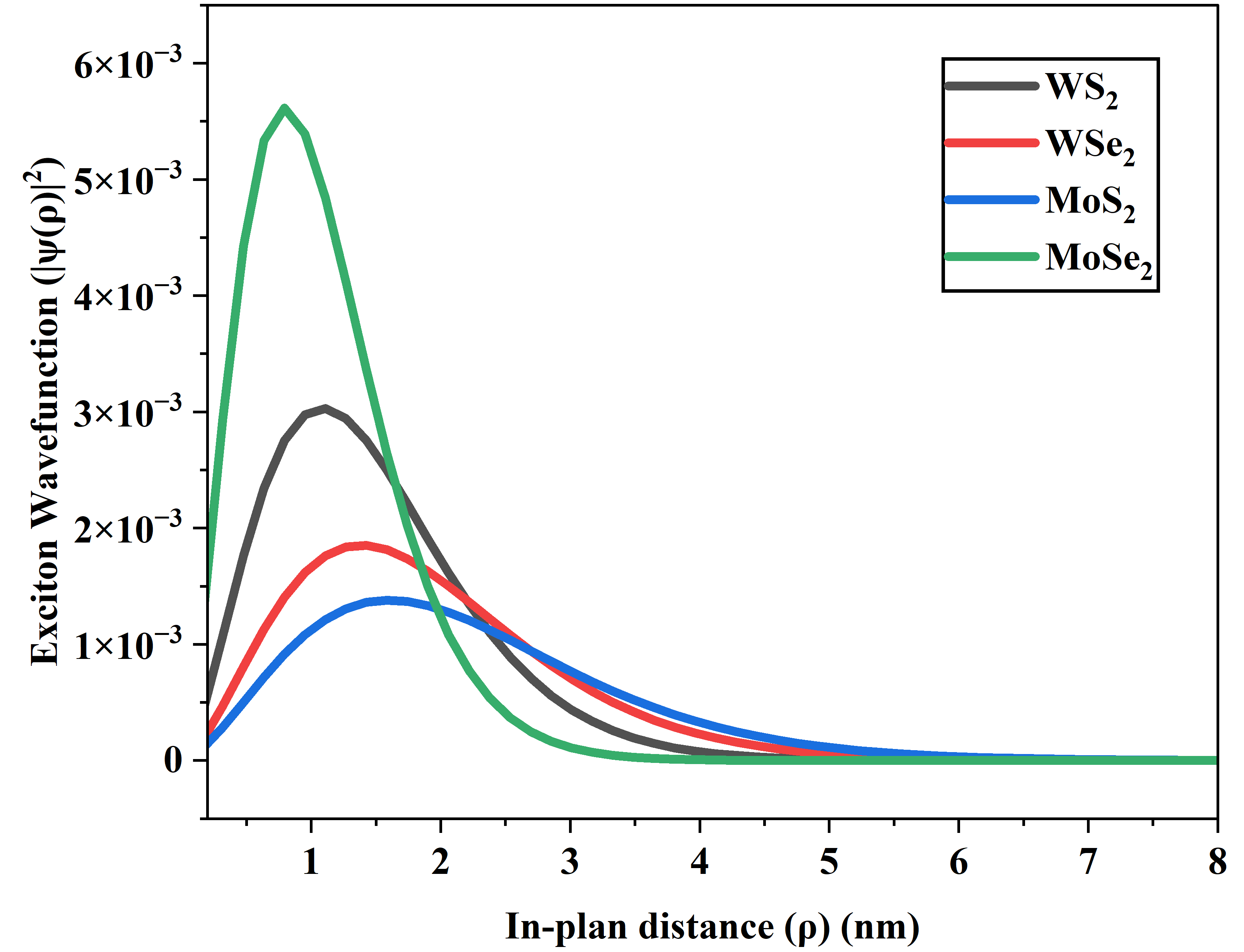}
\par\caption{Exciton states wavefunctions $\psi(\rho)$ for the different monolayer TMDs in the in-plane direction}
\label{fig5}
\end{figure}

\section{Results and Discussion}
\begin{table}
\caption{Band offsets and material properties used in the simulation}
\centering 
\begin{tabular}{c c c c c p{2cm}}

\hline \hline
\text{Monolayer} & \text{Thickness} & \text{$V_e$\cite{deng2018stability}} & \text{$V_h$\cite{deng2018stability}} & \text{$\epsilon_{b}$\cite{laturia2018dielectric}} \\

\text{} & \text{(nm)} & \text{(eV)} & \text{(eV)} & \text{} \\ \hline


\text{$WS_2$} & \text{0.6 \cite{lee2020measurement}} & \text{0.6068} & \text{0.90599} & \text{6.1} \\

\text{$WSe_2$} & \text{0.75 \cite{zhang2020ultrafast}} & \text{0.572649} & \text{0.81197} & \text{7.4} \\

\text{$MoS_2$} & \text{0.65 \cite{li2015two}} & \text{0.2992} & \text{0.7522} & \text{6.2} \\

\text{$MoSe_2$} & \text{0.7  \cite{cowie2021high}} & \text{0.4444} & \text{0.68376} & \text{7.2} \\ \hline

\label{table1}
\end{tabular}
\end{table}
\begin{table*}
\caption{Exciton dynamics for the different monolayer TMDs at room temperature (300K) }
\centering 
\begin{tabular}{c c c c c c c c c c p{2cm}}

\hline \hline
\text{Monolayer} & $m_{e}$ & $m_{h}$ & $E_{g}$ & $E_{X}$ & $E_{X}$ & $E_{b}$ & $E_{b}$ & $d_{cv}$ & $\tau_{\text{eff}}(T)$~\ref{eq17a} & PL Decay~\cite{mohamed2017long} \\

\text{} & \text{} & \text{} & \text{(eV)} & \text{(eV)} & \text{(eV)} & \text{(meV)} & \text{(meV)} & \text{(nm)} & \text{(ns)} & \text{(ns) @300K} \\ \hline

\text{} & \text{} & \text{} & \text{}  & \text{This work} & \text{PL Data} & \text{This work} & \text{PL Data} & & \text{This work} & \text{Average} \\

\text{$WS_2$} & \text{0.33 \cite{conti2020transition}} & \text{0.49 \cite{wu2019exciton}} & \text{1.79 \cite{conti2020transition}} & \text{2.08} & \text{2.027 \cite{sun2023enhanced}} &\text{590} & \text{710$\pm$10\cite{zhu2015exciton}} &  \text{0.6038 \cite{montblanch2021confinement}
} & \text{0.128} & \text{0.22$\pm$7} \\ 

\text{$WSe_2$} & \text{0.48 \cite{rasmussen2015computational}} & \text{0.44 \cite{rasmussen2015computational}} & \text{1.54 \cite{rasmussen2015computational}} & \text{1.68} & \text{1.63 \cite{yan2014photoluminescence}} & \text{536} & \text{370 \cite{he2014tightly}} & \text{0.6038 \cite{montblanch2021confinement}} & \text{0.208} & \text{0.38$\pm$0.13} \\ 

\text{$MoS_2$} & \text{0.47 \cite{wu2019exciton}} & \text{0.54 \cite{wu2019exciton}} & \text{1.72 \cite{deng2018stability}} & \text{1.83} & \text{1.86 \cite{eknapakul2014electronic}} & \text{498} & \text{440 \cite{vaquero2020excitons}} & \text{0.47 \cite{leisgang2020giant}  
} & \text{0.354} & \text{0.42$\pm$0.11} \\ 

\text{$MoSe_2$} & \text{0.43 \cite{conti2020transition}} & \text{0.50 \cite{conti2020transition}} & \text{1.47 \cite{conti2020transition}} & \text{1.64} & \text{1.64 \cite{wang2022disorder}} & \text{469} & \text{550 \cite{ugeda2014giant}} & \text{0.5 \cite{jasinski2025quadrupolar} 
} & \text{0.427} & \text{0.36$\pm$0.1} \\ \hline

\label{table2}
\end{tabular}
\end{table*}

We take the values of parameters effective mass \cite{kylanpaa2015binding}, Band gaps \cite{deng2018stability}, Band$-$offsets \cite{deng2018stability}, and transition dipole moment from the references \cite{wang2019limits,leisgang2020giant, jasinski2025quadrupolar}. The in-plane dielectric constant values are taken from a computational estimation at zero strain \cite{laturia2018dielectric}. The ground-state wavefunctions shown in Fig. \ref{fig5} in the device direction (x) are obtained by solving equation (\ref{eq2a}) and are plotted in Fig.   \ref{fig3}.
The Coulomb potential ($V_C$) for the exciton wavefunction is dependent on the overlap between the electron and hole wavefunctions, derived from equation (\ref{eq8a}). As seen in Fig. \ ref {fig4} $V_C$ profile changes with the TMD material and is the main factor behind the different exciton properties observed.
The associated ground-state Bohr radius $a_B$ is $\sim$1nm, and the wavefunction spans multiple lattice constants (and lattice constant for $WSe_2$ is 0.33 nm) \cite{wang2018colloquium}). 
Therefore, the Wannier-Mott exciton model remains applicable.

The $V_C$ profile varies with the TMD material and correspondingly, the exciton in-plane wavefunctions are plotted in Fig. \ref{fig5}. The exciton energy is estimated 2.08 eV for $WS_2$, while the PL measured around value is 2.016 eV \cite{lee2020measurement} and the obtained binding energy of $WS_2$ is 505 meV, which is closer to the SVM obtained model shown in Fig. \ref{fig8} and Table \ref{table2}. The $WS_2$ and $MoS_2$, having weaker $\epsilon_b$ 6.1 and 6.2, show binding energy close to the SVM model in Fig.  \ref{fig8}.

\begin{figure}[h]
\includegraphics[width=8.5cm]{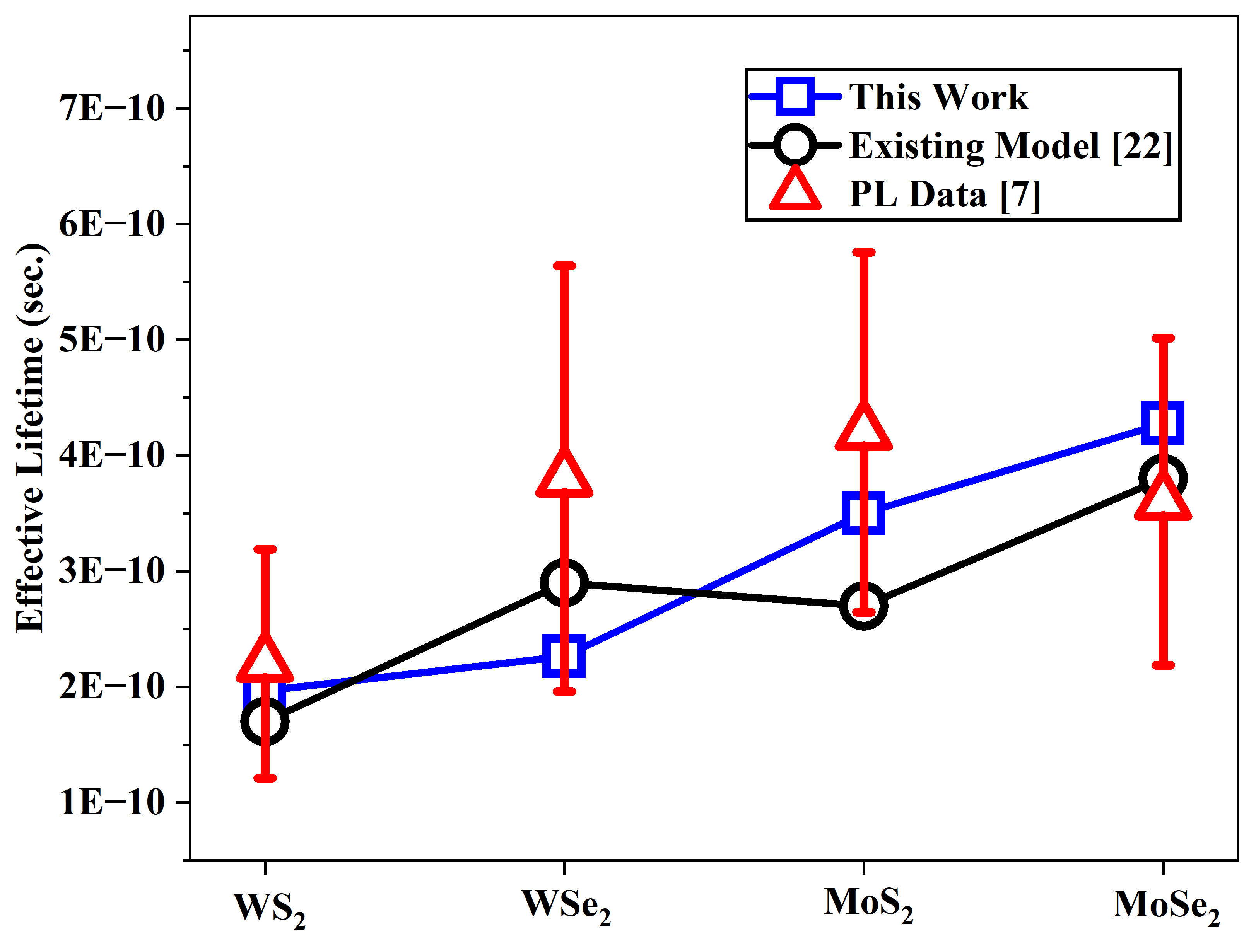}
\par\caption{Comparison of effective lifetime with exisitng model \cite{palummo2015exciton}  and PL measured, shown in table \ref{table2} \cite{mohamed2017long} for monolayer TMDs.}
\label{fig6}
\end{figure}

\begin{figure}[h]
\includegraphics[width=8.5cm]{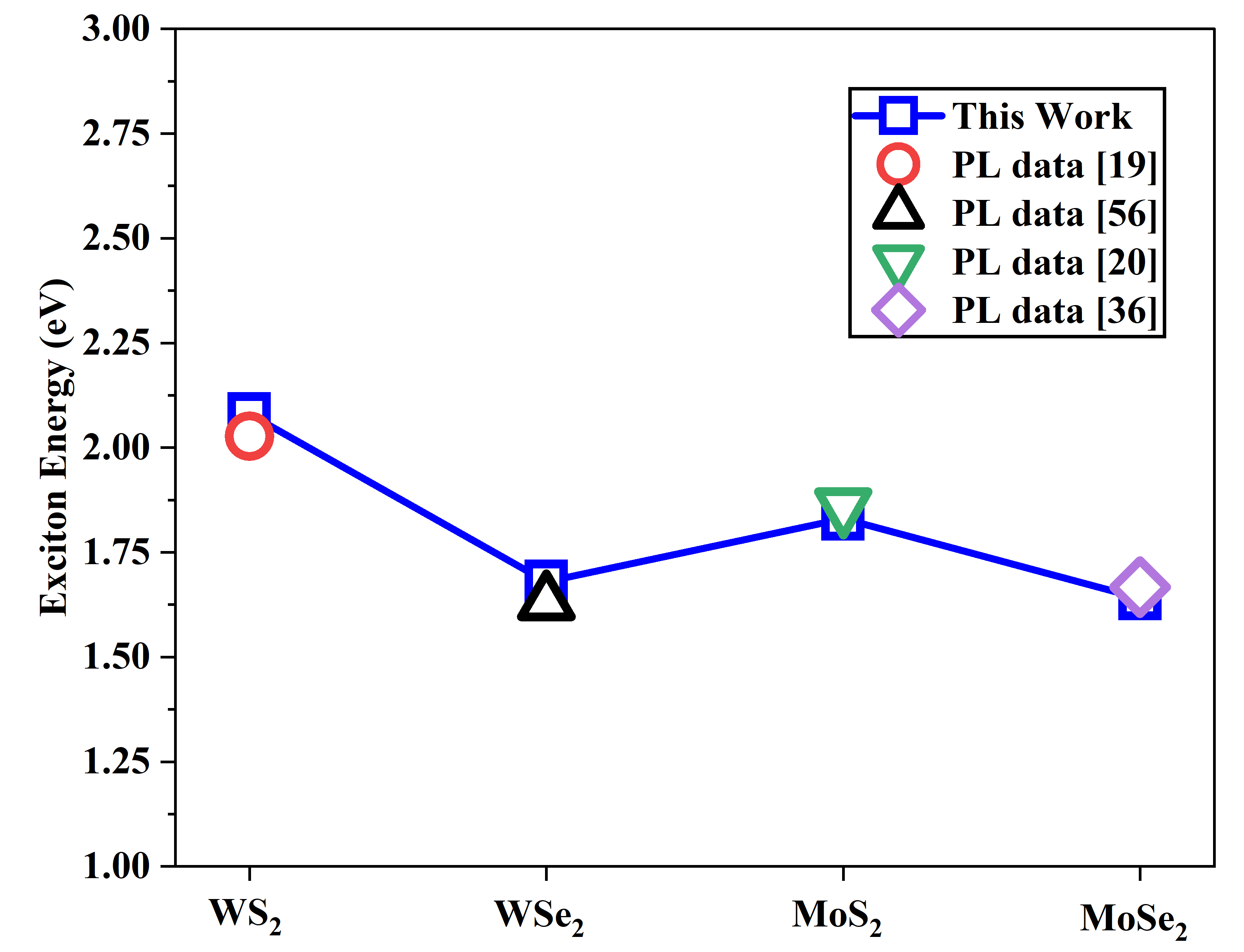}
\par\caption{Comparison of the model$-$estimated Exciton energies with experimentally measured values for different monolayer TMDs \cite{sun2023enhanced,eknapakul2014electronic,wang2022disorder,yan2014photoluminescence}.}
\label{fig7}
\end{figure}

\begin{figure}[h]
\includegraphics[width=8.5cm]{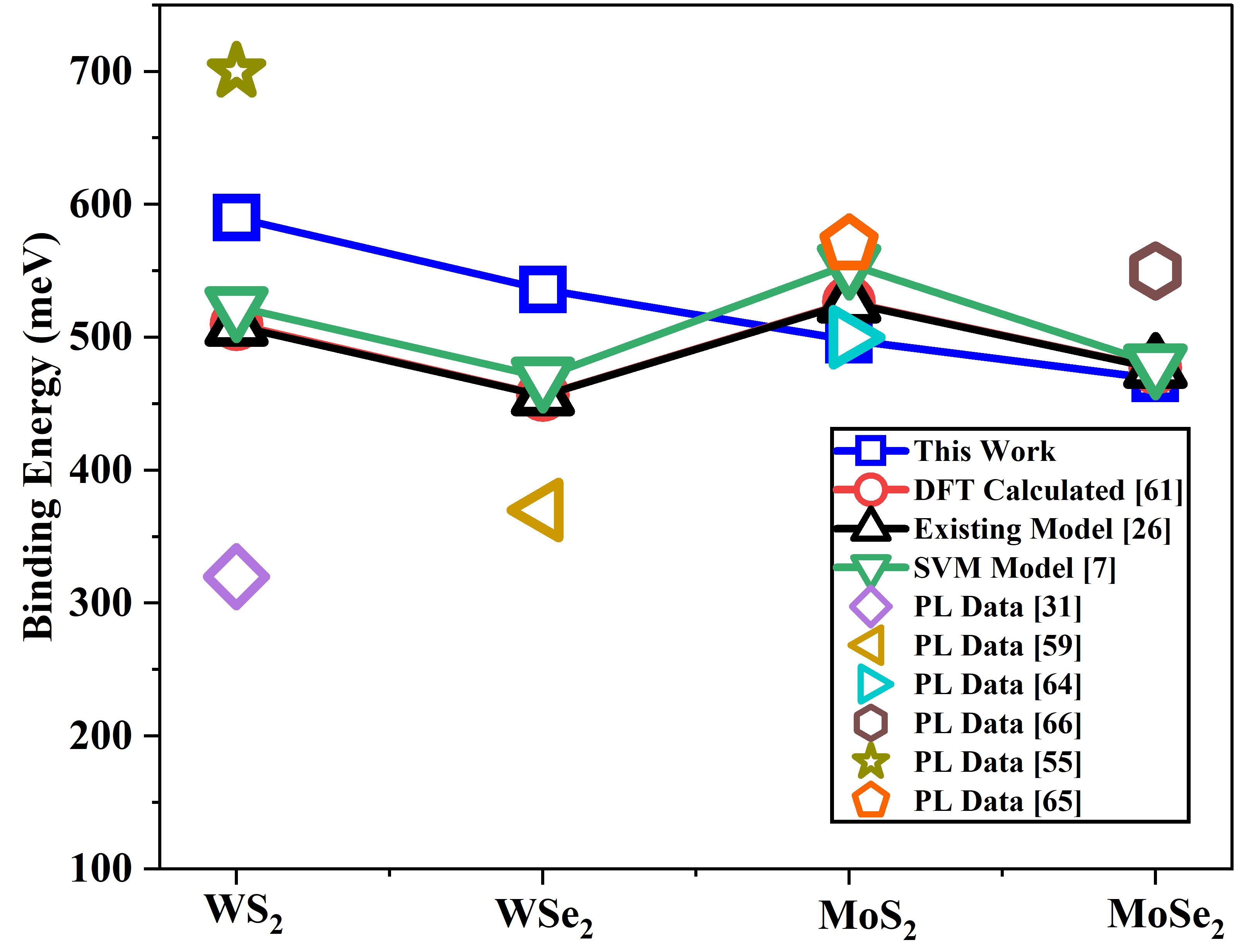}
\par\caption{Comparison of exciton binding energy with DFT calculated \cite{kylanpaa2015binding}, Existing model \cite{wu2019exciton}, SVM model  \cite{zhang2015excited} and PL data \cite{chernikov2014exciton,ye2014probing,he2014tightly,mai2014many,klots2014probing,ugeda2014giant} for the monolayer TMDs.}
\label{fig8}
\end{figure}

\begin{figure}[ht]
\includegraphics[width=8.5cm]{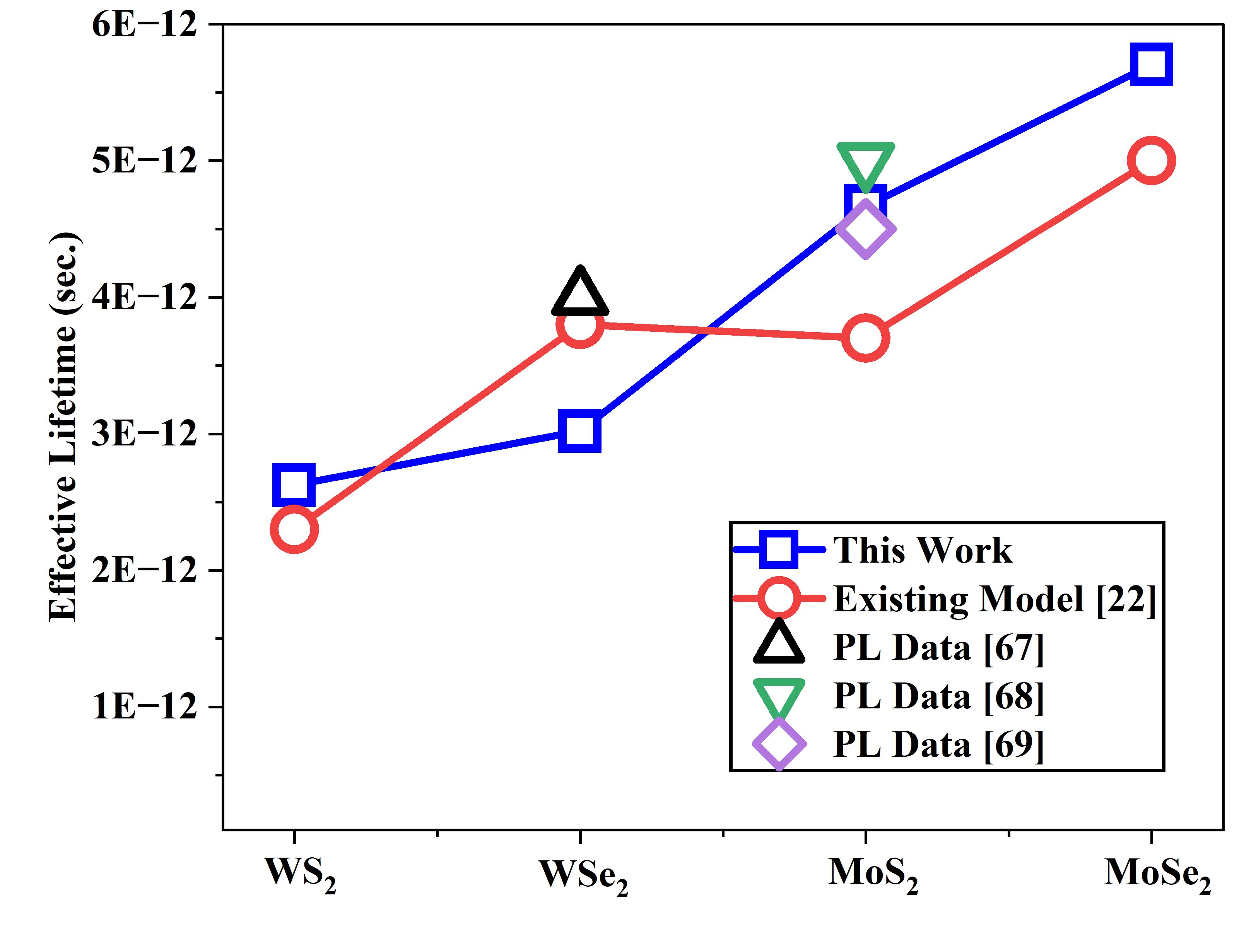}
\par\caption{Low temperature (4K) effective decay lifetime comparison with existing model \cite{palummo2015exciton}, and PL data \cite{wang2014valley,lagarde2014carrier,korn2011low} for monolayer TMDs}
\label{fig9}
\end{figure}

The calculated intrinsic radiative lifetimes are in the range the 128 to 427 picoseconds. 
As shown in Table \ref{table2} and Fig.  \ref{fig6}, the monolayer $MoS_2$ recorded less binding energy compared to other TMDs.
The medium binding energy material $WSe_2$ decays faster compared to $MoS_2$ which can have applications in the light-emitting diode and photodetector. The band offsets of $MoS_2$ and $WS_2$ are higher than those of $WSe_2$ and $MoSe_2$ and leads to higher exciton energy, shown in Fig. \ref{fig7}. 
The effective lifetime of the exciton in monolayer TMDs has been measured at room temperature (300K), and it's significantly longer than the lifetimes observed at low temperatures \cite{palummo2015exciton,mohamed2017long}.
The thermalization of the exciton in monolayer TMDs needs to be considered for the energy redistribution and relaxation processes. After applied photoexcitation begins which creates hot carriers and undergoes rapid electron-hole scattering, followed by exciton formation and cooling via exciton-phonon interactions over a few picoseconds. The subsequent exciton relaxation strongly affect the exciton lifetime, which varies across different TMDs. We observe that the exciton lifetime in monolayers $WS_2$, and $WSe_2$ are shorter (0.15 to 0.26 nanoseconds), and the recombination occurs rapidly. 
In contrast, $MoS_2$ and $MoSe_2$ exhibit longer exciton lifetimes ($\sim$ 0.116 and $\sim$0.199 nanoseconds), allowing more efficient thermalization before recombination, which could be advantageous for applications in single photon emitters \cite{koperski2015single,he2015single}, information storage \cite{qiu2024two}, etc. Additionally, we find that the dielectric constant and transition dipole moment play a significant role in modulating the decay process. \\
At low temperature (at 4K), time-resolved luminescence spectroscopic measurement \cite{palummo2015exciton} shows a lifetime of 2.3 to 5 picoseconds, which nearly matches with the effective lifetime in our calculation as seen in Fig. \ref{fig9}.
DFT calculations for the decay lifetimes for TMDs have been carried out in Ref. \cite{palummo2015exciton} and the effective lifetimes at low (4K) and higher temperatures (300K) are found close to our model predictions as seen in Fig. \ref{fig6}. The EMA approach may not capture all the complexities of the material, like defect states, electronic structure, and dynamic mechanisms, leading to discrepancies between the computed and experimental values. For $MoS_2$, the calculated effective radiation lifetime in this work is smaller than the time-resolved PL measurements \cite{mohamed2017long}. This is due to the contribution of the non-radiative processes and exciton-exciton recombination of Auger type \cite{robert2016exciton}.\\


\vspace{-1em}

In table \ref{table2}, the estimated radiative decay lifetimes for monolayer TMD semiconductors are in the range of 147 to 546 picoseconds, which is purely dependent on the effective mass, band offsets, and dipole matrix element. The model can be extended further for multilayer TMD structures, Interlayer excitons, and also for trions with further modification in the oscillator strength expression. These works we reserve for the near future.

\section{Conclusion}
We propose a mathematical model to calculate the lifetime and exciton binding energy for monolayer TMDs.The dependence of exciton binding energy and lifetime on the properties of a quantum well potential height, thickness, dielectric constant ($\epsilon_b$), and reduced effective mass ($m_r$) (\ref{eq5a}) is brought out.
In this work, we use the quantum transmitting boundary method for the electron and hole wavefunctions in the device direction and a finite difference scheme to calculate the exciton dynamics in the in-plane direction. We compared the calculated values with measurements from photoluminescence and fluorescence at low temperatures and room temperature. The calculated effective radiative lifetime values show a close agreement with the photoluminescence-measured data at low temperatures. A lifetime of thermalized exciton in nanoseconds is also found in agreement with experimentally observed values. The model estimated $MoS_2$ lifetime is higher due to non-radiative recombination of excitons.

\subsection*{\bf Declaration of Competing Interest}
The authors state that they have no known financial interests or personal relationships that could have influenced the work presented in this paper.

 \subsection*{\bf Acknowledgments}
AM gratefully acknowledges support from the Science and Engineering Research Board, Government of India, under Grant CRG/2022/009394.



\bibliography{ref.bib}

\begin{thebibliography}{70}%
\makeatletter
\providecommand \@ifxundefined [1]{%
 \@ifx{#1\undefined}
}%
\providecommand \@ifnum [1]{%
 \ifnum #1\expandafter \@firstoftwo
 \else \expandafter \@secondoftwo
 \fi
}%
\providecommand \@ifx [1]{%
 \ifx #1\expandafter \@firstoftwo
 \else \expandafter \@secondoftwo
 \fi
}%
\providecommand \natexlab [1]{#1}%
\providecommand \enquote  [1]{``#1''}%
\providecommand \bibnamefont  [1]{#1}%
\providecommand \bibfnamefont [1]{#1}%
\providecommand \citenamefont [1]{#1}%
\providecommand \href@noop [0]{\@secondoftwo}%
\providecommand \href [0]{\begingroup \@sanitize@url \@href}%
\providecommand \@href[1]{\@@startlink{#1}\@@href}%
\providecommand \@@href[1]{\endgroup#1\@@endlink}%
\providecommand \@sanitize@url [0]{\catcode `\\12\catcode `\$12\catcode `\&12\catcode `\#12\catcode `\^12\catcode `\_12\catcode `\%12\relax}%
\providecommand \@@startlink[1]{}%
\providecommand \@@endlink[0]{}%
\providecommand \url  [0]{\begingroup\@sanitize@url \@url }%
\providecommand \@url [1]{\endgroup\@href {#1}{\urlprefix }}%
\providecommand \urlprefix  [0]{URL }%
\providecommand \Eprint [0]{\href }%
\providecommand \doibase [0]{http://dx.doi.org/}%
\providecommand \selectlanguage [0]{\@gobble}%
\providecommand \bibinfo  [0]{\@secondoftwo}%
\providecommand \bibfield  [0]{\@secondoftwo}%
\providecommand \translation [1]{[#1]}%
\providecommand \BibitemOpen [0]{}%
\providecommand \bibitemStop [0]{}%
\providecommand \bibitemNoStop [0]{.\EOS\space}%
\providecommand \EOS [0]{\spacefactor3000\relax}%
\providecommand \BibitemShut  [1]{\csname bibitem#1\endcsname}%
\let\auto@bib@innerbib\@empty
\bibitem [{\citenamefont {Mueller}\ and\ \citenamefont {Malic}(2018)}]{mueller2018exciton}%
  \BibitemOpen
  \bibfield  {author} {\bibinfo {author} {\bibfnamefont {T.}~\bibnamefont {Mueller}}\ and\ \bibinfo {author} {\bibfnamefont {E.}~\bibnamefont {Malic}},\ }\bibfield  {title} {\enquote {\bibinfo {title} {Exciton physics and device application of two-dimensional transition metal dichalcogenide semiconductors},}\ }\href {https://doi.org/10.1038/s41699-018-0074-2} {\bibfield  {journal} {\bibinfo  {journal} {npj 2D Materials and Applications}\ }\textbf {\bibinfo {volume} {2}},\ \bibinfo {pages} {29} (\bibinfo {year} {2018})}\BibitemShut {NoStop}%
\bibitem [{\citenamefont {Ahmed}\ and\ \citenamefont {Yi}(2017)}]{ahmed2017two}%
  \BibitemOpen
  \bibfield  {author} {\bibinfo {author} {\bibfnamefont {S.}~\bibnamefont {Ahmed}}\ and\ \bibinfo {author} {\bibfnamefont {J.}~\bibnamefont {Yi}},\ }\bibfield  {title} {\enquote {\bibinfo {title} {Two-dimensional transition metal dichalcogenides and their charge carrier mobilities in field-effect transistors},}\ }\href {https://doi.org/10.1007/s40820-017-0152-6} {\bibfield  {journal} {\bibinfo  {journal} {Nano-micro letters}\ }\textbf {\bibinfo {volume} {9}},\ \bibinfo {pages} {1--23} (\bibinfo {year} {2017})}\BibitemShut {NoStop}%
\bibitem [{\citenamefont {Liu}\ \emph {et~al.}(2015)\citenamefont {Liu}, \citenamefont {Fathi}, \citenamefont {Chen}, \citenamefont {Abbas}, \citenamefont {Ma},\ and\ \citenamefont {Zhou}}]{liu2015chemical}%
  \BibitemOpen
  \bibfield  {author} {\bibinfo {author} {\bibfnamefont {B.}~\bibnamefont {Liu}}, \bibinfo {author} {\bibfnamefont {M.}~\bibnamefont {Fathi}}, \bibinfo {author} {\bibfnamefont {L.}~\bibnamefont {Chen}}, \bibinfo {author} {\bibfnamefont {A.}~\bibnamefont {Abbas}}, \bibinfo {author} {\bibfnamefont {Y.}~\bibnamefont {Ma}}, \ and\ \bibinfo {author} {\bibfnamefont {C.}~\bibnamefont {Zhou}},\ }\bibfield  {title} {\enquote {\bibinfo {title} {Chemical vapor deposition growth of monolayer wse2 with tunable device characteristics and growth mechanism study},}\ }\href {https://doi.org/10.1021/acsnano.5b01301} {\bibfield  {journal} {\bibinfo  {journal} {ACS nano}\ }\textbf {\bibinfo {volume} {9}},\ \bibinfo {pages} {6119--6127} (\bibinfo {year} {2015})}\BibitemShut {NoStop}%
\bibitem [{\citenamefont {Lee}\ \emph {et~al.}(2014)\citenamefont {Lee}, \citenamefont {Lee}, \citenamefont {Van Der~Zande}, \citenamefont {Chen}, \citenamefont {Li}, \citenamefont {Han}, \citenamefont {Cui}, \citenamefont {Arefe}, \citenamefont {Nuckolls}, \citenamefont {Heinz} \emph {et~al.}}]{lee2014atomically}%
  \BibitemOpen
  \bibfield  {author} {\bibinfo {author} {\bibfnamefont {C.-H.}\ \bibnamefont {Lee}}, \bibinfo {author} {\bibfnamefont {G.-H.}\ \bibnamefont {Lee}}, \bibinfo {author} {\bibfnamefont {A.~M.}\ \bibnamefont {Van Der~Zande}}, \bibinfo {author} {\bibfnamefont {W.}~\bibnamefont {Chen}}, \bibinfo {author} {\bibfnamefont {Y.}~\bibnamefont {Li}}, \bibinfo {author} {\bibfnamefont {M.}~\bibnamefont {Han}}, \bibinfo {author} {\bibfnamefont {X.}~\bibnamefont {Cui}}, \bibinfo {author} {\bibfnamefont {G.}~\bibnamefont {Arefe}}, \bibinfo {author} {\bibfnamefont {C.}~\bibnamefont {Nuckolls}}, \bibinfo {author} {\bibfnamefont {T.~F.}\ \bibnamefont {Heinz}},  \emph {et~al.},\ }\bibfield  {title} {\enquote {\bibinfo {title} {Atomically thin p--n junctions with van der waals heterointerfaces},}\ }\href {https://doi.org/10.1038/nnano.2014.150} {\bibfield  {journal} {\bibinfo  {journal} {Nature nanotechnology}\ }\textbf {\bibinfo {volume} {9}},\ \bibinfo {pages} {676--681} (\bibinfo {year} {2014})}\BibitemShut {NoStop}%
\bibitem [{\citenamefont {Baugher}\ \emph {et~al.}()\citenamefont {Baugher}, \citenamefont {Churchill}, \citenamefont {Yang},\ and\ \citenamefont {Jarillo-Herrero}}]{baugheroptoelectronics}%
  \BibitemOpen
  \bibfield  {author} {\bibinfo {author} {\bibfnamefont {B.}~\bibnamefont {Baugher}}, \bibinfo {author} {\bibfnamefont {H.}~\bibnamefont {Churchill}}, \bibinfo {author} {\bibfnamefont {Y.}~\bibnamefont {Yang}}, \ and\ \bibinfo {author} {\bibfnamefont {P.}~\bibnamefont {Jarillo-Herrero}},\ }\bibfield  {title} {\enquote {\bibinfo {title} {Optoelectronics with electrically tunable pn diodes in a monolayer dichalcogenide},}\ }\href {https://doi.org/10.1038/nnano.2014.25} {\bibinfo  {journal} {arXiv preprint arXiv:1310.0452}\ }\BibitemShut {NoStop}%
\bibitem [{\citenamefont {Yin}\ \emph {et~al.}(2012)\citenamefont {Yin}, \citenamefont {Li}, \citenamefont {Li}, \citenamefont {Jiang}, \citenamefont {Shi}, \citenamefont {Sun}, \citenamefont {Lu}, \citenamefont {Zhang}, \citenamefont {Chen},\ and\ \citenamefont {Zhang}}]{yin2012single}%
  \BibitemOpen
\bibfield  {journal} {  }\bibfield  {author} {\bibinfo {author} {\bibfnamefont {Z.}~\bibnamefont {Yin}}, \bibinfo {author} {\bibfnamefont {H.}~\bibnamefont {Li}}, \bibinfo {author} {\bibfnamefont {H.}~\bibnamefont {Li}}, \bibinfo {author} {\bibfnamefont {L.}~\bibnamefont {Jiang}}, \bibinfo {author} {\bibfnamefont {Y.}~\bibnamefont {Shi}}, \bibinfo {author} {\bibfnamefont {Y.}~\bibnamefont {Sun}}, \bibinfo {author} {\bibfnamefont {G.}~\bibnamefont {Lu}}, \bibinfo {author} {\bibfnamefont {Q.}~\bibnamefont {Zhang}}, \bibinfo {author} {\bibfnamefont {X.}~\bibnamefont {Chen}}, \ and\ \bibinfo {author} {\bibfnamefont {H.}~\bibnamefont {Zhang}},\ }\bibfield  {title} {\enquote {\bibinfo {title} {Single-layer mos2 phototransistors},}\ }\href {https://doi.org/10.1021/nn2024557} {\bibfield  {journal} {\bibinfo  {journal} {ACS nano}\ }\textbf {\bibinfo {volume} {6}},\ \bibinfo {pages} {74--80} (\bibinfo {year} {2012})}\BibitemShut {NoStop}%
\bibitem [{\citenamefont {Zhang}, \citenamefont {Kidd},\ and\ \citenamefont {Varga}(2015)}]{zhang2015excited}%
  \BibitemOpen
  \bibfield  {author} {\bibinfo {author} {\bibfnamefont {D.~K.}\ \bibnamefont {Zhang}}, \bibinfo {author} {\bibfnamefont {D.~W.}\ \bibnamefont {Kidd}}, \ and\ \bibinfo {author} {\bibfnamefont {K.}~\bibnamefont {Varga}},\ }\bibfield  {title} {\enquote {\bibinfo {title} {Excited biexcitons in transition metal dichalcogenides},}\ }\href {https://doi.org/10.1021/acs.nanolett.5b03009} {\bibfield  {journal} {\bibinfo  {journal} {Nano letters}\ }\textbf {\bibinfo {volume} {15}},\ \bibinfo {pages} {7002--7005} (\bibinfo {year} {2015})}\BibitemShut {NoStop}%
\bibitem [{\citenamefont {Mohamed}\ \emph {et~al.}(2017)\citenamefont {Mohamed}, \citenamefont {Lim}, \citenamefont {Wang}, \citenamefont {Koirala}, \citenamefont {Mouri}, \citenamefont {Shinokita}, \citenamefont {Miyauchi},\ and\ \citenamefont {Matsuda}}]{mohamed2017long}%
  \BibitemOpen
  \bibfield  {author} {\bibinfo {author} {\bibfnamefont {N.~B.}\ \bibnamefont {Mohamed}}, \bibinfo {author} {\bibfnamefont {H.~E.}\ \bibnamefont {Lim}}, \bibinfo {author} {\bibfnamefont {F.}~\bibnamefont {Wang}}, \bibinfo {author} {\bibfnamefont {S.}~\bibnamefont {Koirala}}, \bibinfo {author} {\bibfnamefont {S.}~\bibnamefont {Mouri}}, \bibinfo {author} {\bibfnamefont {K.}~\bibnamefont {Shinokita}}, \bibinfo {author} {\bibfnamefont {Y.}~\bibnamefont {Miyauchi}}, \ and\ \bibinfo {author} {\bibfnamefont {K.}~\bibnamefont {Matsuda}},\ }\bibfield  {title} {\enquote {\bibinfo {title} {Long radiative lifetimes of excitons in monolayer transition-metal dichalcogenides mx2 (m= mo, w; x= s, se)},}\ }\href {10.7567/APEX.11.015201} {\bibfield  {journal} {\bibinfo  {journal} {Applied Physics Express}\ }\textbf {\bibinfo {volume} {11}},\ \bibinfo {pages} {015201} (\bibinfo {year} {2017})}\BibitemShut {NoStop}%
\bibitem [{\citenamefont {Gong}\ \emph {et~al.}(2013)\citenamefont {Gong}, \citenamefont {Liu}, \citenamefont {Yu}, \citenamefont {Xiao}, \citenamefont {Cui}, \citenamefont {Xu},\ and\ \citenamefont {Yao}}]{gong2013magnetoelectric}%
  \BibitemOpen
  \bibfield  {author} {\bibinfo {author} {\bibfnamefont {Z.}~\bibnamefont {Gong}}, \bibinfo {author} {\bibfnamefont {G.-B.}\ \bibnamefont {Liu}}, \bibinfo {author} {\bibfnamefont {H.}~\bibnamefont {Yu}}, \bibinfo {author} {\bibfnamefont {D.}~\bibnamefont {Xiao}}, \bibinfo {author} {\bibfnamefont {X.}~\bibnamefont {Cui}}, \bibinfo {author} {\bibfnamefont {X.}~\bibnamefont {Xu}}, \ and\ \bibinfo {author} {\bibfnamefont {W.}~\bibnamefont {Yao}},\ }\bibfield  {title} {\enquote {\bibinfo {title} {Magnetoelectric effects and valley-controlled spin quantum gates in transition metal dichalcogenide bilayers},}\ }\href {https://doi.org/10.1038/nnano.2014.25} {\bibfield  {journal} {\bibinfo  {journal} {Nature communications}\ }\textbf {\bibinfo {volume} {4}},\ \bibinfo {pages} {2053} (\bibinfo {year} {2013})}\BibitemShut {NoStop}%
\bibitem [{\citenamefont {Jiang}\ \emph {et~al.}(2018)\citenamefont {Jiang}, \citenamefont {Xu}, \citenamefont {Rasmita}, \citenamefont {Huang}, \citenamefont {Li}, \citenamefont {Xiong},\ and\ \citenamefont {Gao}}]{jiang2018microsecond}%
  \BibitemOpen
  \bibfield  {author} {\bibinfo {author} {\bibfnamefont {C.}~\bibnamefont {Jiang}}, \bibinfo {author} {\bibfnamefont {W.}~\bibnamefont {Xu}}, \bibinfo {author} {\bibfnamefont {A.}~\bibnamefont {Rasmita}}, \bibinfo {author} {\bibfnamefont {Z.}~\bibnamefont {Huang}}, \bibinfo {author} {\bibfnamefont {K.}~\bibnamefont {Li}}, \bibinfo {author} {\bibfnamefont {Q.}~\bibnamefont {Xiong}}, \ and\ \bibinfo {author} {\bibfnamefont {W.-b.}\ \bibnamefont {Gao}},\ }\bibfield  {title} {\enquote {\bibinfo {title} {Microsecond dark-exciton valley polarization memory in two-dimensional heterostructures},}\ }\href {https://doi.org/10.1038/s41467-018-03174-3} {\bibfield  {journal} {\bibinfo  {journal} {Nature communications}\ }\textbf {\bibinfo {volume} {9}},\ \bibinfo {pages} {753} (\bibinfo {year} {2018})}\BibitemShut {NoStop}%
\bibitem [{\citenamefont {Liu}\ and\ \citenamefont {Hersam}(2019)}]{liu20192d}%
  \BibitemOpen
  \bibfield  {author} {\bibinfo {author} {\bibfnamefont {X.}~\bibnamefont {Liu}}\ and\ \bibinfo {author} {\bibfnamefont {M.~C.}\ \bibnamefont {Hersam}},\ }\bibfield  {title} {\enquote {\bibinfo {title} {2d materials for quantum information science},}\ }\href {https://doi.org/10.1038/s41578-019-0136-x} {\bibfield  {journal} {\bibinfo  {journal} {Nature Reviews Materials}\ }\textbf {\bibinfo {volume} {4}},\ \bibinfo {pages} {669--684} (\bibinfo {year} {2019})}\BibitemShut {NoStop}%
\bibitem [{\citenamefont {Gan}, \citenamefont {Zhao},\ and\ \citenamefont {Quan}(2017)}]{gan2017two}%
  \BibitemOpen
  \bibfield  {author} {\bibinfo {author} {\bibfnamefont {X.}~\bibnamefont {Gan}}, \bibinfo {author} {\bibfnamefont {H.}~\bibnamefont {Zhao}}, \ and\ \bibinfo {author} {\bibfnamefont {X.}~\bibnamefont {Quan}},\ }\bibfield  {title} {\enquote {\bibinfo {title} {Two-dimensional mos2: A promising building block for biosensors},}\ }\href {https://doi.org/10.1016/j.bios.2016.03.042} {\bibfield  {journal} {\bibinfo  {journal} {Biosensors and Bioelectronics}\ }\textbf {\bibinfo {volume} {89}},\ \bibinfo {pages} {56--71} (\bibinfo {year} {2017})}\BibitemShut {NoStop}%
\bibitem [{\citenamefont {Lopez-Sanchez}\ \emph {et~al.}(2013)\citenamefont {Lopez-Sanchez}, \citenamefont {Lembke}, \citenamefont {Kayci}, \citenamefont {Radenovic},\ and\ \citenamefont {Kis}}]{lopez2013ultrasensitive}%
  \BibitemOpen
  \bibfield  {author} {\bibinfo {author} {\bibfnamefont {O.}~\bibnamefont {Lopez-Sanchez}}, \bibinfo {author} {\bibfnamefont {D.}~\bibnamefont {Lembke}}, \bibinfo {author} {\bibfnamefont {M.}~\bibnamefont {Kayci}}, \bibinfo {author} {\bibfnamefont {A.}~\bibnamefont {Radenovic}}, \ and\ \bibinfo {author} {\bibfnamefont {A.}~\bibnamefont {Kis}},\ }\bibfield  {title} {\enquote {\bibinfo {title} {Ultrasensitive photodetectors based on monolayer mos2},}\ }\href {https://doi.org/10.1038/nnano.2013.100} {\bibfield  {journal} {\bibinfo  {journal} {Nature nanotechnology}\ }\textbf {\bibinfo {volume} {8}},\ \bibinfo {pages} {497--501} (\bibinfo {year} {2013})}\BibitemShut {NoStop}%
\bibitem [{\citenamefont {Lukman}\ \emph {et~al.}(2020)\citenamefont {Lukman}, \citenamefont {Ding}, \citenamefont {Xu}, \citenamefont {Tao}, \citenamefont {Riis-Jensen}, \citenamefont {Zhang}, \citenamefont {Wu}, \citenamefont {Yang}, \citenamefont {Luo}, \citenamefont {Hsu} \emph {et~al.}}]{lukman2020high}%
  \BibitemOpen
  \bibfield  {author} {\bibinfo {author} {\bibfnamefont {S.}~\bibnamefont {Lukman}}, \bibinfo {author} {\bibfnamefont {L.}~\bibnamefont {Ding}}, \bibinfo {author} {\bibfnamefont {L.}~\bibnamefont {Xu}}, \bibinfo {author} {\bibfnamefont {Y.}~\bibnamefont {Tao}}, \bibinfo {author} {\bibfnamefont {A.~C.}\ \bibnamefont {Riis-Jensen}}, \bibinfo {author} {\bibfnamefont {G.}~\bibnamefont {Zhang}}, \bibinfo {author} {\bibfnamefont {Q.~Y.~S.}\ \bibnamefont {Wu}}, \bibinfo {author} {\bibfnamefont {M.}~\bibnamefont {Yang}}, \bibinfo {author} {\bibfnamefont {S.}~\bibnamefont {Luo}}, \bibinfo {author} {\bibfnamefont {C.}~\bibnamefont {Hsu}},  \emph {et~al.},\ }\bibfield  {title} {\enquote {\bibinfo {title} {High oscillator strength interlayer excitons in two-dimensional heterostructures for mid-infrared photodetection},}\ }\href {https://doi.org/10.1038/s41565-020-0717-2} {\bibfield  {journal} {\bibinfo  {journal} {Nature nanotechnology}\ }\textbf {\bibinfo {volume} {15}},\ \bibinfo {pages} {675--682} (\bibinfo {year}
  {2020})}\BibitemShut {NoStop}%
\bibitem [{\citenamefont {Ross}\ \emph {et~al.}(2014)\citenamefont {Ross}, \citenamefont {Klement}, \citenamefont {Jones}, \citenamefont {Ghimire}, \citenamefont {Yan}, \citenamefont {Mandrus}, \citenamefont {Taniguchi}, \citenamefont {Watanabe}, \citenamefont {Kitamura}, \citenamefont {Yao} \emph {et~al.}}]{ross2014electrically}%
  \BibitemOpen
  \bibfield  {author} {\bibinfo {author} {\bibfnamefont {J.~S.}\ \bibnamefont {Ross}}, \bibinfo {author} {\bibfnamefont {P.}~\bibnamefont {Klement}}, \bibinfo {author} {\bibfnamefont {A.~M.}\ \bibnamefont {Jones}}, \bibinfo {author} {\bibfnamefont {N.~J.}\ \bibnamefont {Ghimire}}, \bibinfo {author} {\bibfnamefont {J.}~\bibnamefont {Yan}}, \bibinfo {author} {\bibfnamefont {D.}~\bibnamefont {Mandrus}}, \bibinfo {author} {\bibfnamefont {T.}~\bibnamefont {Taniguchi}}, \bibinfo {author} {\bibfnamefont {K.}~\bibnamefont {Watanabe}}, \bibinfo {author} {\bibfnamefont {K.}~\bibnamefont {Kitamura}}, \bibinfo {author} {\bibfnamefont {W.}~\bibnamefont {Yao}},  \emph {et~al.},\ }\bibfield  {title} {\enquote {\bibinfo {title} {Electrically tunable excitonic light-emitting diodes based on monolayer wse2 p--n junctions},}\ }\href {https://doi.org/10.1038/nnano.2014.26} {\bibfield  {journal} {\bibinfo  {journal} {Nature nanotechnology}\ }\textbf {\bibinfo {volume} {9}},\ \bibinfo {pages} {268--272} (\bibinfo {year}
  {2014})}\BibitemShut {NoStop}%
\bibitem [{\citenamefont {Pospischil}, \citenamefont {Furchi},\ and\ \citenamefont {Mueller}(2014)}]{pospischil2014solar}%
  \BibitemOpen
  \bibfield  {author} {\bibinfo {author} {\bibfnamefont {A.}~\bibnamefont {Pospischil}}, \bibinfo {author} {\bibfnamefont {M.~M.}\ \bibnamefont {Furchi}}, \ and\ \bibinfo {author} {\bibfnamefont {T.}~\bibnamefont {Mueller}},\ }\bibfield  {title} {\enquote {\bibinfo {title} {Solar-energy conversion and light emission in an atomic monolayer p--n diode},}\ }\href {https://doi.org/10.1038/nnano.2014.14} {\bibfield  {journal} {\bibinfo  {journal} {Nature nanotechnology}\ }\textbf {\bibinfo {volume} {9}},\ \bibinfo {pages} {257--261} (\bibinfo {year} {2014})}\BibitemShut {NoStop}%
\bibitem [{\citenamefont {Koperski}\ \emph {et~al.}(2015)\citenamefont {Koperski}, \citenamefont {Nogajewski}, \citenamefont {Arora}, \citenamefont {Cherkez}, \citenamefont {Mallet}, \citenamefont {Veuillen}, \citenamefont {Marcus}, \citenamefont {Kossacki},\ and\ \citenamefont {Potemski}}]{koperski2015single}%
  \BibitemOpen
  \bibfield  {author} {\bibinfo {author} {\bibfnamefont {M.}~\bibnamefont {Koperski}}, \bibinfo {author} {\bibfnamefont {K.}~\bibnamefont {Nogajewski}}, \bibinfo {author} {\bibfnamefont {A.}~\bibnamefont {Arora}}, \bibinfo {author} {\bibfnamefont {V.}~\bibnamefont {Cherkez}}, \bibinfo {author} {\bibfnamefont {P.}~\bibnamefont {Mallet}}, \bibinfo {author} {\bibfnamefont {J.-Y.}\ \bibnamefont {Veuillen}}, \bibinfo {author} {\bibfnamefont {J.}~\bibnamefont {Marcus}}, \bibinfo {author} {\bibfnamefont {P.}~\bibnamefont {Kossacki}}, \ and\ \bibinfo {author} {\bibfnamefont {M.}~\bibnamefont {Potemski}},\ }\bibfield  {title} {\enquote {\bibinfo {title} {Single photon emitters in exfoliated wse2 structures},}\ }\href {https://doi.org/10.1038/nnano.2015.67} {\bibfield  {journal} {\bibinfo  {journal} {Nature nanotechnology}\ }\textbf {\bibinfo {volume} {10}},\ \bibinfo {pages} {503--506} (\bibinfo {year} {2015})}\BibitemShut {NoStop}%
\bibitem [{\citenamefont {He}\ \emph {et~al.}(2015)\citenamefont {He}, \citenamefont {Clark}, \citenamefont {Schaibley}, \citenamefont {He}, \citenamefont {Chen}, \citenamefont {Wei}, \citenamefont {Ding}, \citenamefont {Zhang}, \citenamefont {Yao}, \citenamefont {Xu} \emph {et~al.}}]{he2015single}%
  \BibitemOpen
  \bibfield  {author} {\bibinfo {author} {\bibfnamefont {Y.-M.}\ \bibnamefont {He}}, \bibinfo {author} {\bibfnamefont {G.}~\bibnamefont {Clark}}, \bibinfo {author} {\bibfnamefont {J.~R.}\ \bibnamefont {Schaibley}}, \bibinfo {author} {\bibfnamefont {Y.}~\bibnamefont {He}}, \bibinfo {author} {\bibfnamefont {M.-C.}\ \bibnamefont {Chen}}, \bibinfo {author} {\bibfnamefont {Y.-J.}\ \bibnamefont {Wei}}, \bibinfo {author} {\bibfnamefont {X.}~\bibnamefont {Ding}}, \bibinfo {author} {\bibfnamefont {Q.}~\bibnamefont {Zhang}}, \bibinfo {author} {\bibfnamefont {W.}~\bibnamefont {Yao}}, \bibinfo {author} {\bibfnamefont {X.}~\bibnamefont {Xu}},  \emph {et~al.},\ }\bibfield  {title} {\enquote {\bibinfo {title} {Single quantum emitters in monolayer semiconductors},}\ }\href {https://doi.org/10.1038/nnano.2015.75} {\bibfield  {journal} {\bibinfo  {journal} {Nature nanotechnology}\ }\textbf {\bibinfo {volume} {10}},\ \bibinfo {pages} {497--502} (\bibinfo {year} {2015})}\BibitemShut {NoStop}%
\bibitem [{\citenamefont {Sun}, \citenamefont {Lu},\ and\ \citenamefont {Lu}(2023)}]{sun2023enhanced}%
  \BibitemOpen
  \bibfield  {author} {\bibinfo {author} {\bibfnamefont {X.}~\bibnamefont {Sun}}, \bibinfo {author} {\bibfnamefont {Z.}~\bibnamefont {Lu}}, \ and\ \bibinfo {author} {\bibfnamefont {Y.}~\bibnamefont {Lu}},\ }\bibfield  {title} {\enquote {\bibinfo {title} {Enhanced interactions of excitonic complexes in free-standing ws 2},}\ }\href@noop {} {\bibfield  {journal} {\bibinfo  {journal} {Nanoscale}\ }\textbf {\bibinfo {volume} {15}},\ \bibinfo {pages} {19533--19545} (\bibinfo {year} {2023})}\BibitemShut {NoStop}%
\bibitem [{\citenamefont {Eknapakul}\ \emph {et~al.}(2014)\citenamefont {Eknapakul}, \citenamefont {King}, \citenamefont {Asakawa}, \citenamefont {Buaphet}, \citenamefont {He}, \citenamefont {Mo}, \citenamefont {Takagi}, \citenamefont {Shen}, \citenamefont {Baumberger}, \citenamefont {Sasagawa} \emph {et~al.}}]{eknapakul2014electronic}%
  \BibitemOpen
  \bibfield  {author} {\bibinfo {author} {\bibfnamefont {T.}~\bibnamefont {Eknapakul}}, \bibinfo {author} {\bibfnamefont {P.}~\bibnamefont {King}}, \bibinfo {author} {\bibfnamefont {M.}~\bibnamefont {Asakawa}}, \bibinfo {author} {\bibfnamefont {P.}~\bibnamefont {Buaphet}}, \bibinfo {author} {\bibfnamefont {R.-H.}\ \bibnamefont {He}}, \bibinfo {author} {\bibfnamefont {S.-K.}\ \bibnamefont {Mo}}, \bibinfo {author} {\bibfnamefont {H.}~\bibnamefont {Takagi}}, \bibinfo {author} {\bibfnamefont {K.}~\bibnamefont {Shen}}, \bibinfo {author} {\bibfnamefont {F.}~\bibnamefont {Baumberger}}, \bibinfo {author} {\bibfnamefont {T.}~\bibnamefont {Sasagawa}},  \emph {et~al.},\ }\bibfield  {title} {\enquote {\bibinfo {title} {Electronic structure of a quasi-freestanding mos2 monolayer},}\ }\href@noop {} {\bibfield  {journal} {\bibinfo  {journal} {Nano letters}\ }\textbf {\bibinfo {volume} {14}},\ \bibinfo {pages} {1312--1316} (\bibinfo {year} {2014})}\BibitemShut {NoStop}%
\bibitem [{\citenamefont {Qiu}\ \emph {et~al.}(2024)\citenamefont {Qiu}, \citenamefont {Yu}, \citenamefont {Zhao}, \citenamefont {Zhang}, \citenamefont {Xu}, \citenamefont {Li}, \citenamefont {Li}, \citenamefont {Bao}, \citenamefont {Chai}, \citenamefont {Chen} \emph {et~al.}}]{qiu2024two}%
  \BibitemOpen
  \bibfield  {author} {\bibinfo {author} {\bibfnamefont {H.}~\bibnamefont {Qiu}}, \bibinfo {author} {\bibfnamefont {Z.}~\bibnamefont {Yu}}, \bibinfo {author} {\bibfnamefont {T.}~\bibnamefont {Zhao}}, \bibinfo {author} {\bibfnamefont {Q.}~\bibnamefont {Zhang}}, \bibinfo {author} {\bibfnamefont {M.}~\bibnamefont {Xu}}, \bibinfo {author} {\bibfnamefont {P.}~\bibnamefont {Li}}, \bibinfo {author} {\bibfnamefont {T.}~\bibnamefont {Li}}, \bibinfo {author} {\bibfnamefont {W.}~\bibnamefont {Bao}}, \bibinfo {author} {\bibfnamefont {Y.}~\bibnamefont {Chai}}, \bibinfo {author} {\bibfnamefont {S.}~\bibnamefont {Chen}},  \emph {et~al.},\ }\bibfield  {title} {\enquote {\bibinfo {title} {Two-dimensional materials for future information technology: status and prospects},}\ }\href {https://doi.org/10.1007/s11432-024-4033-8} {\bibfield  {journal} {\bibinfo  {journal} {Science China Information Sciences}\ }\textbf {\bibinfo {volume} {67}},\ \bibinfo {pages} {1--147} (\bibinfo {year} {2024})}\BibitemShut {NoStop}%
\bibitem [{\citenamefont {Deng}, \citenamefont {Li},\ and\ \citenamefont {Li}(2018)}]{deng2018stability}%
  \BibitemOpen
  \bibfield  {author} {\bibinfo {author} {\bibfnamefont {S.}~\bibnamefont {Deng}}, \bibinfo {author} {\bibfnamefont {L.}~\bibnamefont {Li}}, \ and\ \bibinfo {author} {\bibfnamefont {M.}~\bibnamefont {Li}},\ }\bibfield  {title} {\enquote {\bibinfo {title} {Stability of direct band gap under mechanical strains for monolayer mos2, mose2, ws2 and wse2},}\ }\href {https://doi.org/10.1016/j.physe.2018.03.016} {\bibfield  {journal} {\bibinfo  {journal} {Physica E: Low-dimensional Systems and Nanostructures}\ }\textbf {\bibinfo {volume} {101}},\ \bibinfo {pages} {44--49} (\bibinfo {year} {2018})}\BibitemShut {NoStop}%
\bibitem [{\citenamefont {Palummo}, \citenamefont {Bernardi},\ and\ \citenamefont {Grossman}(2015)}]{palummo2015exciton}%
  \BibitemOpen
  \bibfield  {author} {\bibinfo {author} {\bibfnamefont {M.}~\bibnamefont {Palummo}}, \bibinfo {author} {\bibfnamefont {M.}~\bibnamefont {Bernardi}}, \ and\ \bibinfo {author} {\bibfnamefont {J.~C.}\ \bibnamefont {Grossman}},\ }\bibfield  {title} {\enquote {\bibinfo {title} {Exciton radiative lifetimes in two-dimensional transition metal dichalcogenides},}\ }\href {https://doi.org/10.1021/nl503799t} {\bibfield  {journal} {\bibinfo  {journal} {Nano letters}\ }\textbf {\bibinfo {volume} {15}},\ \bibinfo {pages} {2794--2800} (\bibinfo {year} {2015})}\BibitemShut {NoStop}%
\bibitem [{\citenamefont {Wang}\ \emph {et~al.}(2016)\citenamefont {Wang}, \citenamefont {Zhang}, \citenamefont {Chan}, \citenamefont {Manolatou}, \citenamefont {Tiwari},\ and\ \citenamefont {Rana}}]{wang2016radiative}%
  \BibitemOpen
  \bibfield  {author} {\bibinfo {author} {\bibfnamefont {H.}~\bibnamefont {Wang}}, \bibinfo {author} {\bibfnamefont {C.}~\bibnamefont {Zhang}}, \bibinfo {author} {\bibfnamefont {W.}~\bibnamefont {Chan}}, \bibinfo {author} {\bibfnamefont {C.}~\bibnamefont {Manolatou}}, \bibinfo {author} {\bibfnamefont {S.}~\bibnamefont {Tiwari}}, \ and\ \bibinfo {author} {\bibfnamefont {F.}~\bibnamefont {Rana}},\ }\bibfield  {title} {\enquote {\bibinfo {title} {Radiative lifetimes of excitons and trions in monolayers of the metal dichalcogenide mos 2},}\ }\href {https://doi.org/10.1103/PhysRevB.93.045407} {\bibfield  {journal} {\bibinfo  {journal} {Physical Review B}\ }\textbf {\bibinfo {volume} {93}},\ \bibinfo {pages} {045407} (\bibinfo {year} {2016})}\BibitemShut {NoStop}%
\bibitem [{\citenamefont {Bergh{\"a}user}\ and\ \citenamefont {Malic}(2014)}]{berghauser2014analytical}%
  \BibitemOpen
  \bibfield  {author} {\bibinfo {author} {\bibfnamefont {G.}~\bibnamefont {Bergh{\"a}user}}\ and\ \bibinfo {author} {\bibfnamefont {E.}~\bibnamefont {Malic}},\ }\bibfield  {title} {\enquote {\bibinfo {title} {Analytical approach to excitonic properties of mos 2},}\ }\href {https://doi.org/10.1103/PhysRevB.89.125309} {\bibfield  {journal} {\bibinfo  {journal} {Physical Review B}\ }\textbf {\bibinfo {volume} {89}},\ \bibinfo {pages} {125309} (\bibinfo {year} {2014})}\BibitemShut {NoStop}%
\bibitem [{\citenamefont {Cheiwchanchamnangij}\ and\ \citenamefont {Lambrecht}(2012)}]{cheiwchanchamnangij2012quasiparticle}%
  \BibitemOpen
  \bibfield  {author} {\bibinfo {author} {\bibfnamefont {T.}~\bibnamefont {Cheiwchanchamnangij}}\ and\ \bibinfo {author} {\bibfnamefont {W.~R.}\ \bibnamefont {Lambrecht}},\ }\bibfield  {title} {\enquote {\bibinfo {title} {Quasiparticle band structure calculation of monolayer, bilayer, and bulk mos 2},}\ }\href {https://doi.org/10.1103/PhysRevB.85.205302} {\bibfield  {journal} {\bibinfo  {journal} {Physical Review B—Condensed Matter and Materials Physics}\ }\textbf {\bibinfo {volume} {85}},\ \bibinfo {pages} {205302} (\bibinfo {year} {2012})}\BibitemShut {NoStop}%
\bibitem [{\citenamefont {Wu}, \citenamefont {Cheng},\ and\ \citenamefont {Wang}(2019)}]{wu2019exciton}%
  \BibitemOpen
  \bibfield  {author} {\bibinfo {author} {\bibfnamefont {S.}~\bibnamefont {Wu}}, \bibinfo {author} {\bibfnamefont {L.}~\bibnamefont {Cheng}}, \ and\ \bibinfo {author} {\bibfnamefont {Q.}~\bibnamefont {Wang}},\ }\bibfield  {title} {\enquote {\bibinfo {title} {Exciton states and absorption spectra in freestanding monolayer transition metal dichalcogenides: A variationally optimized diagonalization method},}\ }\href {https://doi.org/10.1103/PhysRevB.100.115430} {\bibfield  {journal} {\bibinfo  {journal} {Physical Review B}\ }\textbf {\bibinfo {volume} {100}},\ \bibinfo {pages} {115430} (\bibinfo {year} {2019})}\BibitemShut {NoStop}%
\bibitem [{\citenamefont {Martins~Quintela}\ and\ \citenamefont {Peres}(2020)}]{martins2020colloquium}%
  \BibitemOpen
  \bibfield  {author} {\bibinfo {author} {\bibfnamefont {M.~F.}\ \bibnamefont {Martins~Quintela}}\ and\ \bibinfo {author} {\bibfnamefont {N.~M.}\ \bibnamefont {Peres}},\ }\bibfield  {title} {\enquote {\bibinfo {title} {A colloquium on the variational method applied to excitons in 2d materials},}\ }\href {https://doi.org/10.1140/epjb/e2020-10490-9} {\bibfield  {journal} {\bibinfo  {journal} {The European Physical Journal B}\ }\textbf {\bibinfo {volume} {93}},\ \bibinfo {pages} {1--16} (\bibinfo {year} {2020})}\BibitemShut {NoStop}%
\bibitem [{\citenamefont {Kronik}\ and\ \citenamefont {Neaton}(2016)}]{kronik2016excited}%
  \BibitemOpen
  \bibfield  {author} {\bibinfo {author} {\bibfnamefont {L.}~\bibnamefont {Kronik}}\ and\ \bibinfo {author} {\bibfnamefont {J.~B.}\ \bibnamefont {Neaton}},\ }\bibfield  {title} {\enquote {\bibinfo {title} {Excited-state properties of molecular solids from first principles},}\ }\href {https://doi.org/10.1146/annurev-physchem-040214-121351} {\bibfield  {journal} {\bibinfo  {journal} {Annual review of physical chemistry}\ }\textbf {\bibinfo {volume} {67}},\ \bibinfo {pages} {587--616} (\bibinfo {year} {2016})}\BibitemShut {NoStop}%
\bibitem [{\citenamefont {Lee}\ \emph {et~al.}(2020)\citenamefont {Lee}, \citenamefont {Seo}, \citenamefont {Kwon}, \citenamefont {Kim}, \citenamefont {Kim}, \citenamefont {Yun}, \citenamefont {Cha}, \citenamefont {Song}, \citenamefont {Lee}, \citenamefont {Jung} \emph {et~al.}}]{lee2020measurement}%
  \BibitemOpen
  \bibfield  {author} {\bibinfo {author} {\bibfnamefont {M.-J.}\ \bibnamefont {Lee}}, \bibinfo {author} {\bibfnamefont {D.~H.}\ \bibnamefont {Seo}}, \bibinfo {author} {\bibfnamefont {S.~M.}\ \bibnamefont {Kwon}}, \bibinfo {author} {\bibfnamefont {D.}~\bibnamefont {Kim}}, \bibinfo {author} {\bibfnamefont {Y.}~\bibnamefont {Kim}}, \bibinfo {author} {\bibfnamefont {W.~S.}\ \bibnamefont {Yun}}, \bibinfo {author} {\bibfnamefont {J.-H.}\ \bibnamefont {Cha}}, \bibinfo {author} {\bibfnamefont {H.-K.}\ \bibnamefont {Song}}, \bibinfo {author} {\bibfnamefont {S.}~\bibnamefont {Lee}}, \bibinfo {author} {\bibfnamefont {M.}~\bibnamefont {Jung}},  \emph {et~al.},\ }\bibfield  {title} {\enquote {\bibinfo {title} {Measurement of exciton and trion energies in multistacked hbn/ws2 coupled quantum wells for resonant tunneling diodes},}\ }\href {https://doi.org/10.1021/acsnano.0c08133} {\bibfield  {journal} {\bibinfo  {journal} {ACS nano}\ }\textbf {\bibinfo {volume} {14}},\ \bibinfo {pages} {16114--16121} (\bibinfo {year}
  {2020})}\BibitemShut {NoStop}%
\bibitem [{\citenamefont {Yuan}\ and\ \citenamefont {Huang}(2015)}]{yuan2015exciton}%
  \BibitemOpen
  \bibfield  {author} {\bibinfo {author} {\bibfnamefont {L.}~\bibnamefont {Yuan}}\ and\ \bibinfo {author} {\bibfnamefont {L.}~\bibnamefont {Huang}},\ }\bibfield  {title} {\enquote {\bibinfo {title} {Exciton dynamics and annihilation in ws 2 2d semiconductors},}\ }\href {https://doi.org/10.1039/C5NR00383K} {\bibfield  {journal} {\bibinfo  {journal} {Nanoscale}\ }\textbf {\bibinfo {volume} {7}},\ \bibinfo {pages} {7402--7408} (\bibinfo {year} {2015})}\BibitemShut {NoStop}%
\bibitem [{\citenamefont {Zhu}, \citenamefont {Chen},\ and\ \citenamefont {Cui}(2015)}]{zhu2015exciton}%
  \BibitemOpen
  \bibfield  {author} {\bibinfo {author} {\bibfnamefont {B.}~\bibnamefont {Zhu}}, \bibinfo {author} {\bibfnamefont {X.}~\bibnamefont {Chen}}, \ and\ \bibinfo {author} {\bibfnamefont {X.}~\bibnamefont {Cui}},\ }\bibfield  {title} {\enquote {\bibinfo {title} {Exciton binding energy of monolayer ws2},}\ }\href {https://doi.org/10.1007/s40820-017-0152-6} {\bibfield  {journal} {\bibinfo  {journal} {Scientific reports}\ }\textbf {\bibinfo {volume} {5}},\ \bibinfo {pages} {9218} (\bibinfo {year} {2015})}\BibitemShut {NoStop}%
\bibitem [{\citenamefont {Chernikov}\ \emph {et~al.}(2014)\citenamefont {Chernikov}, \citenamefont {Berkelbach}, \citenamefont {Hill}, \citenamefont {Rigosi}, \citenamefont {Li}, \citenamefont {Aslan}, \citenamefont {Reichman}, \citenamefont {Hybertsen},\ and\ \citenamefont {Heinz}}]{chernikov2014exciton}%
  \BibitemOpen
  \bibfield  {author} {\bibinfo {author} {\bibfnamefont {A.}~\bibnamefont {Chernikov}}, \bibinfo {author} {\bibfnamefont {T.~C.}\ \bibnamefont {Berkelbach}}, \bibinfo {author} {\bibfnamefont {H.~M.}\ \bibnamefont {Hill}}, \bibinfo {author} {\bibfnamefont {A.}~\bibnamefont {Rigosi}}, \bibinfo {author} {\bibfnamefont {Y.}~\bibnamefont {Li}}, \bibinfo {author} {\bibfnamefont {B.}~\bibnamefont {Aslan}}, \bibinfo {author} {\bibfnamefont {D.~R.}\ \bibnamefont {Reichman}}, \bibinfo {author} {\bibfnamefont {M.~S.}\ \bibnamefont {Hybertsen}}, \ and\ \bibinfo {author} {\bibfnamefont {T.~F.}\ \bibnamefont {Heinz}},\ }\bibfield  {title} {\enquote {\bibinfo {title} {Exciton binding energy and nonhydrogenic rydberg series in monolayer ws 2},}\ }\href {https://doi.org/10.1103/PhysRevLett.113.076802} {\bibfield  {journal} {\bibinfo  {journal} {Physical review letters}\ }\textbf {\bibinfo {volume} {113}},\ \bibinfo {pages} {076802} (\bibinfo {year} {2014})}\BibitemShut {NoStop}%
\bibitem [{\citenamefont {Chernikov}\ \emph {et~al.}(2015)\citenamefont {Chernikov}, \citenamefont {Van Der~Zande}, \citenamefont {Hill}, \citenamefont {Rigosi}, \citenamefont {Velauthapillai}, \citenamefont {Hone},\ and\ \citenamefont {Heinz}}]{chernikov2015electrical}%
  \BibitemOpen
  \bibfield  {author} {\bibinfo {author} {\bibfnamefont {A.}~\bibnamefont {Chernikov}}, \bibinfo {author} {\bibfnamefont {A.~M.}\ \bibnamefont {Van Der~Zande}}, \bibinfo {author} {\bibfnamefont {H.~M.}\ \bibnamefont {Hill}}, \bibinfo {author} {\bibfnamefont {A.~F.}\ \bibnamefont {Rigosi}}, \bibinfo {author} {\bibfnamefont {A.}~\bibnamefont {Velauthapillai}}, \bibinfo {author} {\bibfnamefont {J.}~\bibnamefont {Hone}}, \ and\ \bibinfo {author} {\bibfnamefont {T.~F.}\ \bibnamefont {Heinz}},\ }\bibfield  {title} {\enquote {\bibinfo {title} {Electrical tuning of exciton binding energies in monolayer ws 2},}\ }\href {https://doi.org/10.1103/PhysRevLett.115.126802} {\bibfield  {journal} {\bibinfo  {journal} {Physical review letters}\ }\textbf {\bibinfo {volume} {115}},\ \bibinfo {pages} {126802} (\bibinfo {year} {2015})}\BibitemShut {NoStop}%
\bibitem [{\citenamefont {Lin}\ \emph {et~al.}(2014)\citenamefont {Lin}, \citenamefont {Ling}, \citenamefont {Yu}, \citenamefont {Huang}, \citenamefont {Hsu}, \citenamefont {Lee}, \citenamefont {Kong}, \citenamefont {Dresselhaus},\ and\ \citenamefont {Palacios}}]{lin2014dielectric}%
  \BibitemOpen
  \bibfield  {author} {\bibinfo {author} {\bibfnamefont {Y.}~\bibnamefont {Lin}}, \bibinfo {author} {\bibfnamefont {X.}~\bibnamefont {Ling}}, \bibinfo {author} {\bibfnamefont {L.}~\bibnamefont {Yu}}, \bibinfo {author} {\bibfnamefont {S.}~\bibnamefont {Huang}}, \bibinfo {author} {\bibfnamefont {A.~L.}\ \bibnamefont {Hsu}}, \bibinfo {author} {\bibfnamefont {Y.-H.}\ \bibnamefont {Lee}}, \bibinfo {author} {\bibfnamefont {J.}~\bibnamefont {Kong}}, \bibinfo {author} {\bibfnamefont {M.~S.}\ \bibnamefont {Dresselhaus}}, \ and\ \bibinfo {author} {\bibfnamefont {T.}~\bibnamefont {Palacios}},\ }\bibfield  {title} {\enquote {\bibinfo {title} {Dielectric screening of excitons and trions in single-layer mos2},}\ }\href {https://doi.org/10.1021/nl501988y} {\bibfield  {journal} {\bibinfo  {journal} {Nano letters}\ }\textbf {\bibinfo {volume} {14}},\ \bibinfo {pages} {5569--5576} (\bibinfo {year} {2014})}\BibitemShut {NoStop}%
\bibitem [{\citenamefont {Wang}\ \emph {et~al.}(2022)\citenamefont {Wang}, \citenamefont {Manolatou}, \citenamefont {Bai}, \citenamefont {Hone}, \citenamefont {Rana},\ and\ \citenamefont {Zhu}}]{wang2022disorder}%
  \BibitemOpen
  \bibfield  {author} {\bibinfo {author} {\bibfnamefont {J.}~\bibnamefont {Wang}}, \bibinfo {author} {\bibfnamefont {C.}~\bibnamefont {Manolatou}}, \bibinfo {author} {\bibfnamefont {Y.}~\bibnamefont {Bai}}, \bibinfo {author} {\bibfnamefont {J.}~\bibnamefont {Hone}}, \bibinfo {author} {\bibfnamefont {F.}~\bibnamefont {Rana}}, \ and\ \bibinfo {author} {\bibfnamefont {X.-Y.}\ \bibnamefont {Zhu}},\ }\bibfield  {title} {\enquote {\bibinfo {title} {Disorder of excitons and trions in monolayer mose2},}\ }\href {https://doi.org/10.1063/5.0108001} {\bibfield  {journal} {\bibinfo  {journal} {The Journal of Chemical Physics}\ }\textbf {\bibinfo {volume} {157}} (\bibinfo {year} {2022})}\BibitemShut {NoStop}%
\bibitem [{\citenamefont {Nimje}\ and\ \citenamefont {Mahajan}(2023)}]{pscripta2023analytical}%
  \BibitemOpen
  \bibfield  {author} {\bibinfo {author} {\bibfnamefont {R.~R.}\ \bibnamefont {Nimje}}\ and\ \bibinfo {author} {\bibfnamefont {A.}~\bibnamefont {Mahajan}},\ }\bibfield  {title} {\enquote {\bibinfo {title} {Analytical estimation of lifetime of quasi-bound states in iii-v semiconductors quantum well},}\ }\href {https://iopscience.iop.org/article/10.1088/1402-4896/acea03/meta} {\bibfield  {journal} {\bibinfo  {journal} {Physica Scripta}\ }\textbf {\bibinfo {volume} {98}},\ \bibinfo {pages} {095014} (\bibinfo {year} {2023})}\BibitemShut {NoStop}%
\bibitem [{\citenamefont {Mahajan}\ and\ \citenamefont {Ganguly}(2021)}]{mahajan2021analytical}%
  \BibitemOpen
  \bibfield  {author} {\bibinfo {author} {\bibfnamefont {A.}~\bibnamefont {Mahajan}}\ and\ \bibinfo {author} {\bibfnamefont {S.}~\bibnamefont {Ganguly}},\ }\bibfield  {title} {\enquote {\bibinfo {title} {An analytical model for electron tunneling in triangular quantum wells},}\ }\href {https://iopscience.iop.org/article/10.1088/1361-6641/abec15/meta} {\bibfield  {journal} {\bibinfo  {journal} {Semiconductor Science and Technology}\ }\textbf {\bibinfo {volume} {36}},\ \bibinfo {pages} {055012} (\bibinfo {year} {2021})}\BibitemShut {NoStop}%
\bibitem [{\citenamefont {Belov}(2019)}]{belov2019energy}%
  \BibitemOpen
  \bibfield  {author} {\bibinfo {author} {\bibfnamefont {P.}~\bibnamefont {Belov}},\ }\bibfield  {title} {\enquote {\bibinfo {title} {Energy spectrum of excitons in square quantum wells},}\ }\href {https://doi.org/10.1016/j.physe.2019.04.008} {\bibfield  {journal} {\bibinfo  {journal} {Physica E: Low-dimensional systems and nanostructures}\ }\textbf {\bibinfo {volume} {112}},\ \bibinfo {pages} {96--108} (\bibinfo {year} {2019})}\BibitemShut {NoStop}%
\bibitem [{\citenamefont {Laturia}, \citenamefont {Van~de Put},\ and\ \citenamefont {Vandenberghe}(2018)}]{laturia2018dielectric}%
  \BibitemOpen
  \bibfield  {author} {\bibinfo {author} {\bibfnamefont {A.}~\bibnamefont {Laturia}}, \bibinfo {author} {\bibfnamefont {M.~L.}\ \bibnamefont {Van~de Put}}, \ and\ \bibinfo {author} {\bibfnamefont {W.~G.}\ \bibnamefont {Vandenberghe}},\ }\bibfield  {title} {\enquote {\bibinfo {title} {Dielectric properties of hexagonal boron nitride and transition metal dichalcogenides: from monolayer to bulk},}\ }\href {https://doi.org/10.1038/s41699-018-0050-x} {\bibfield  {journal} {\bibinfo  {journal} {npj 2D Materials and Applications}\ }\textbf {\bibinfo {volume} {2}},\ \bibinfo {pages} {6} (\bibinfo {year} {2018})}\BibitemShut {NoStop}%
\bibitem [{\citenamefont {Burstein}\ and\ \citenamefont {Weisbuch}(2012)}]{burstein2012confined}%
  \BibitemOpen
  \bibfield  {author} {\bibinfo {author} {\bibfnamefont {E.}~\bibnamefont {Burstein}}\ and\ \bibinfo {author} {\bibfnamefont {C.}~\bibnamefont {Weisbuch}},\ }\href {https://doi.org/10.1007/978-1-4615-1963-8} {\emph {\bibinfo {title} {Confined electrons and photons: New physics and applications}}},\ Vol.\ \bibinfo {volume} {340}\ (\bibinfo  {publisher} {Springer Science \& Business Media},\ \bibinfo {year} {2012})\BibitemShut {NoStop}%
\bibitem [{\citenamefont {Dimmock}(1967)}]{dimmock1967introduction}%
  \BibitemOpen
  \bibfield  {author} {\bibinfo {author} {\bibfnamefont {J.~O.}\ \bibnamefont {Dimmock}},\ }\bibfield  {title} {\enquote {\bibinfo {title} {Introduction to the theory of exciton states in semiconductors},}\ }in\ \href {https://doi.org/10.1016/S0080-8784(08)60319-1} {\emph {\bibinfo {booktitle} {Semiconductors and semimetals}}},\ Vol.~\bibinfo {volume} {3}\ (\bibinfo  {publisher} {Elsevier},\ \bibinfo {year} {1967})\ pp.\ \bibinfo {pages} {259--319}\BibitemShut {NoStop}%
\bibitem [{\citenamefont {Citrin}(1993)}]{citrin1993radiative}%
  \BibitemOpen
  \bibfield  {author} {\bibinfo {author} {\bibfnamefont {D.}~\bibnamefont {Citrin}},\ }\bibfield  {title} {\enquote {\bibinfo {title} {Radiative lifetimes of excitons in quantum wells: Localization and phase-coherence effects},}\ }\href {https://doi.org/10.1103/PhysRevB.47.3832} {\bibfield  {journal} {\bibinfo  {journal} {Physical Review B}\ }\textbf {\bibinfo {volume} {47}},\ \bibinfo {pages} {3832} (\bibinfo {year} {1993})}\BibitemShut {NoStop}%
\bibitem [{\citenamefont {Robert}\ \emph {et~al.}(2016)\citenamefont {Robert}, \citenamefont {Lagarde}, \citenamefont {Cadiz}, \citenamefont {Wang}, \citenamefont {Lassagne}, \citenamefont {Amand}, \citenamefont {Balocchi}, \citenamefont {Renucci}, \citenamefont {Tongay}, \citenamefont {Urbaszek} \emph {et~al.}}]{robert2016exciton}%
  \BibitemOpen
  \bibfield  {author} {\bibinfo {author} {\bibfnamefont {C.}~\bibnamefont {Robert}}, \bibinfo {author} {\bibfnamefont {D.}~\bibnamefont {Lagarde}}, \bibinfo {author} {\bibfnamefont {F.}~\bibnamefont {Cadiz}}, \bibinfo {author} {\bibfnamefont {G.}~\bibnamefont {Wang}}, \bibinfo {author} {\bibfnamefont {B.}~\bibnamefont {Lassagne}}, \bibinfo {author} {\bibfnamefont {T.}~\bibnamefont {Amand}}, \bibinfo {author} {\bibfnamefont {A.}~\bibnamefont {Balocchi}}, \bibinfo {author} {\bibfnamefont {P.}~\bibnamefont {Renucci}}, \bibinfo {author} {\bibfnamefont {S.}~\bibnamefont {Tongay}}, \bibinfo {author} {\bibfnamefont {B.}~\bibnamefont {Urbaszek}},  \emph {et~al.},\ }\bibfield  {title} {\enquote {\bibinfo {title} {Exciton radiative lifetime in transition metal dichalcogenide monolayers},}\ }\href {https://doi.org/10.1103/PhysRevB.93.205423} {\bibfield  {journal} {\bibinfo  {journal} {Physical review B}\ }\textbf {\bibinfo {volume} {93}},\ \bibinfo {pages} {205423} (\bibinfo {year} {2016})}\BibitemShut {NoStop}%
\bibitem [{\citenamefont {Rowlinson*}(2005)}]{rowlinson2005maxwell}%
  \BibitemOpen
  \bibfield  {author} {\bibinfo {author} {\bibfnamefont {J.}~\bibnamefont {Rowlinson*}},\ }\bibfield  {title} {\enquote {\bibinfo {title} {The maxwell--boltzmann distribution},}\ }\href@noop {} {\bibfield  {journal} {\bibinfo  {journal} {Molecular Physics}\ }\textbf {\bibinfo {volume} {103}},\ \bibinfo {pages} {2821--2828} (\bibinfo {year} {2005})}\BibitemShut {NoStop}%
\bibitem [{\citenamefont {Lent}\ and\ \citenamefont {Kirkner}(1990)}]{lent1990quantum}%
  \BibitemOpen
  \bibfield  {author} {\bibinfo {author} {\bibfnamefont {C.~S.}\ \bibnamefont {Lent}}\ and\ \bibinfo {author} {\bibfnamefont {D.~J.}\ \bibnamefont {Kirkner}},\ }\bibfield  {title} {\enquote {\bibinfo {title} {The quantum transmitting boundary method},}\ }\href {https://doi.org/10.1063/1.345156} {\bibfield  {journal} {\bibinfo  {journal} {Journal of Applied Physics}\ }\textbf {\bibinfo {volume} {67}},\ \bibinfo {pages} {6353--6359} (\bibinfo {year} {1990})}\BibitemShut {NoStop}%
\bibitem [{\citenamefont {Shao}\ \emph {et~al.}(1995)\citenamefont {Shao}, \citenamefont {Porod}, \citenamefont {Lent},\ and\ \citenamefont {Kirkner}}]{shao1995eigenvalue}%
  \BibitemOpen
  \bibfield  {author} {\bibinfo {author} {\bibfnamefont {Z.-a.}\ \bibnamefont {Shao}}, \bibinfo {author} {\bibfnamefont {W.}~\bibnamefont {Porod}}, \bibinfo {author} {\bibfnamefont {C.~S.}\ \bibnamefont {Lent}}, \ and\ \bibinfo {author} {\bibfnamefont {D.~J.}\ \bibnamefont {Kirkner}},\ }\bibfield  {title} {\enquote {\bibinfo {title} {An eigenvalue method for open-boundary quantum transmission problems},}\ }\href {https://doi.org/10.1063/1.360132} {\bibfield  {journal} {\bibinfo  {journal} {Journal of applied physics}\ }\textbf {\bibinfo {volume} {78}},\ \bibinfo {pages} {2177--2186} (\bibinfo {year} {1995})}\BibitemShut {NoStop}%
\bibitem [{\citenamefont {Butov}\ \emph {et~al.}(1999)\citenamefont {Butov}, \citenamefont {Imamoglu}, \citenamefont {Mintsev}, \citenamefont {Campman},\ and\ \citenamefont {Gossard}}]{butov1999photoluminescence}%
  \BibitemOpen
  \bibfield  {author} {\bibinfo {author} {\bibfnamefont {L.}~\bibnamefont {Butov}}, \bibinfo {author} {\bibfnamefont {A.}~\bibnamefont {Imamoglu}}, \bibinfo {author} {\bibfnamefont {A.}~\bibnamefont {Mintsev}}, \bibinfo {author} {\bibfnamefont {K.}~\bibnamefont {Campman}}, \ and\ \bibinfo {author} {\bibfnamefont {A.}~\bibnamefont {Gossard}},\ }\bibfield  {title} {\enquote {\bibinfo {title} {Photoluminescence kinetics of indirect excitons in gaas/alxga1-xas coupled quantum wells},}\ }\href {https://doi.org/10.1103/PhysRevB.59.1625} {\bibfield  {journal} {\bibinfo  {journal} {Physical Review B}\ }\textbf {\bibinfo {volume} {59}},\ \bibinfo {pages} {1625} (\bibinfo {year} {1999})}\BibitemShut {NoStop}%
\bibitem [{\citenamefont {Sivalertporn}\ \emph {et~al.}(2012)\citenamefont {Sivalertporn}, \citenamefont {Mouchliadis}, \citenamefont {Ivanov}, \citenamefont {Philp},\ and\ \citenamefont {Muljarov}}]{sivalertporn2012direct}%
  \BibitemOpen
  \bibfield  {author} {\bibinfo {author} {\bibfnamefont {K.}~\bibnamefont {Sivalertporn}}, \bibinfo {author} {\bibfnamefont {L.}~\bibnamefont {Mouchliadis}}, \bibinfo {author} {\bibfnamefont {A.}~\bibnamefont {Ivanov}}, \bibinfo {author} {\bibfnamefont {R.}~\bibnamefont {Philp}}, \ and\ \bibinfo {author} {\bibfnamefont {E.~A.}\ \bibnamefont {Muljarov}},\ }\bibfield  {title} {\enquote {\bibinfo {title} {Direct and indirect excitons in semiconductor coupled quantum wells in an applied electric field},}\ }\href {https://doi.org/10.1103/PhysRevB.85.045207} {\bibfield  {journal} {\bibinfo  {journal} {Physical Review B}\ }\textbf {\bibinfo {volume} {85}},\ \bibinfo {pages} {045207} (\bibinfo {year} {2012})}\BibitemShut {NoStop}%
\bibitem [{\citenamefont {Zhang}\ \emph {et~al.}(2020)\citenamefont {Zhang}, \citenamefont {Chen}, \citenamefont {Yang}, \citenamefont {Liu}, \citenamefont {Ma}, \citenamefont {Li}, \citenamefont {Zhao}, \citenamefont {Luo}, \citenamefont {Duan},\ and\ \citenamefont {Duan}}]{zhang2020ultrafast}%
  \BibitemOpen
  \bibfield  {author} {\bibinfo {author} {\bibfnamefont {Z.}~\bibnamefont {Zhang}}, \bibinfo {author} {\bibfnamefont {P.}~\bibnamefont {Chen}}, \bibinfo {author} {\bibfnamefont {X.}~\bibnamefont {Yang}}, \bibinfo {author} {\bibfnamefont {Y.}~\bibnamefont {Liu}}, \bibinfo {author} {\bibfnamefont {H.}~\bibnamefont {Ma}}, \bibinfo {author} {\bibfnamefont {J.}~\bibnamefont {Li}}, \bibinfo {author} {\bibfnamefont {B.}~\bibnamefont {Zhao}}, \bibinfo {author} {\bibfnamefont {J.}~\bibnamefont {Luo}}, \bibinfo {author} {\bibfnamefont {X.}~\bibnamefont {Duan}}, \ and\ \bibinfo {author} {\bibfnamefont {X.}~\bibnamefont {Duan}},\ }\bibfield  {title} {\enquote {\bibinfo {title} {Ultrafast growth of large single crystals of monolayer ws2 and wse2},}\ }\href {https://doi.org/10.1093/nsr/nwz223} {\bibfield  {journal} {\bibinfo  {journal} {National Science Review}\ }\textbf {\bibinfo {volume} {7}},\ \bibinfo {pages} {737--744} (\bibinfo {year} {2020})}\BibitemShut {NoStop}%
\bibitem [{\citenamefont {Li}\ and\ \citenamefont {Zhu}(2015)}]{li2015two}%
  \BibitemOpen
  \bibfield  {author} {\bibinfo {author} {\bibfnamefont {X.}~\bibnamefont {Li}}\ and\ \bibinfo {author} {\bibfnamefont {H.}~\bibnamefont {Zhu}},\ }\bibfield  {title} {\enquote {\bibinfo {title} {Two-dimensional mos2: Properties, preparation, and applications},}\ }\href {https://doi.org/10.1016/j.jmat.2015.03.003} {\bibfield  {journal} {\bibinfo  {journal} {Journal of Materiomics}\ }\textbf {\bibinfo {volume} {1}},\ \bibinfo {pages} {33--44} (\bibinfo {year} {2015})}\BibitemShut {NoStop}%
\bibitem [{\citenamefont {Cowie}\ \emph {et~al.}(2021)\citenamefont {Cowie}, \citenamefont {Plougmann}, \citenamefont {Benkirane}, \citenamefont {Schu{\'e}}, \citenamefont {Schumacher},\ and\ \citenamefont {Gr{\"u}tter}}]{cowie2021high}%
  \BibitemOpen
  \bibfield  {author} {\bibinfo {author} {\bibfnamefont {M.}~\bibnamefont {Cowie}}, \bibinfo {author} {\bibfnamefont {R.}~\bibnamefont {Plougmann}}, \bibinfo {author} {\bibfnamefont {Y.}~\bibnamefont {Benkirane}}, \bibinfo {author} {\bibfnamefont {L.}~\bibnamefont {Schu{\'e}}}, \bibinfo {author} {\bibfnamefont {Z.}~\bibnamefont {Schumacher}}, \ and\ \bibinfo {author} {\bibfnamefont {P.}~\bibnamefont {Gr{\"u}tter}},\ }\bibfield  {title} {\enquote {\bibinfo {title} {How high is a mose2 monolayer?}}\ }\href {https://iopscience.iop.org/article/10.1088/1361-6528/ac40bd/meta} {\bibfield  {journal} {\bibinfo  {journal} {Nanotechnology}\ }\textbf {\bibinfo {volume} {33}},\ \bibinfo {pages} {125706} (\bibinfo {year} {2021})}\BibitemShut {NoStop}%
\bibitem [{\citenamefont {Conti}\ \emph {et~al.}(2020)\citenamefont {Conti}, \citenamefont {Neilson}, \citenamefont {Peeters},\ and\ \citenamefont {Perali}}]{conti2020transition}%
  \BibitemOpen
  \bibfield  {author} {\bibinfo {author} {\bibfnamefont {S.}~\bibnamefont {Conti}}, \bibinfo {author} {\bibfnamefont {D.}~\bibnamefont {Neilson}}, \bibinfo {author} {\bibfnamefont {F.~M.}\ \bibnamefont {Peeters}}, \ and\ \bibinfo {author} {\bibfnamefont {A.}~\bibnamefont {Perali}},\ }\bibfield  {title} {\enquote {\bibinfo {title} {Transition metal dichalcogenides as strategy for high temperature electron-hole superfluidity},}\ }\href {https://doi.org/10.3390/condmat5010022} {\bibfield  {journal} {\bibinfo  {journal} {Condensed Matter}\ }\textbf {\bibinfo {volume} {5}},\ \bibinfo {pages} {22} (\bibinfo {year} {2020})}\BibitemShut {NoStop}%
\bibitem [{\citenamefont {Montblanch}\ \emph {et~al.}(2021)\citenamefont {Montblanch}, \citenamefont {Kara}, \citenamefont {Paradisanos}, \citenamefont {Purser}, \citenamefont {Feuer}, \citenamefont {Alexeev}, \citenamefont {Stefan}, \citenamefont {Qin}, \citenamefont {Blei}, \citenamefont {Wang} \emph {et~al.}}]{montblanch2021confinement}%
  \BibitemOpen
  \bibfield  {author} {\bibinfo {author} {\bibfnamefont {A.~R.-P.}\ \bibnamefont {Montblanch}}, \bibinfo {author} {\bibfnamefont {D.~M.}\ \bibnamefont {Kara}}, \bibinfo {author} {\bibfnamefont {I.}~\bibnamefont {Paradisanos}}, \bibinfo {author} {\bibfnamefont {C.~M.}\ \bibnamefont {Purser}}, \bibinfo {author} {\bibfnamefont {M.~S.}\ \bibnamefont {Feuer}}, \bibinfo {author} {\bibfnamefont {E.~M.}\ \bibnamefont {Alexeev}}, \bibinfo {author} {\bibfnamefont {L.}~\bibnamefont {Stefan}}, \bibinfo {author} {\bibfnamefont {Y.}~\bibnamefont {Qin}}, \bibinfo {author} {\bibfnamefont {M.}~\bibnamefont {Blei}}, \bibinfo {author} {\bibfnamefont {G.}~\bibnamefont {Wang}},  \emph {et~al.},\ }\bibfield  {title} {\enquote {\bibinfo {title} {Confinement of long-lived interlayer excitons in ws2/wse2 heterostructures},}\ }\href {https://doi.org/10.1038/s42005-021-00625-0} {\bibfield  {journal} {\bibinfo  {journal} {Communications Physics}\ }\textbf {\bibinfo {volume} {4}},\ \bibinfo {pages} {119} (\bibinfo {year}
  {2021})}\BibitemShut {NoStop}%
\bibitem [{\citenamefont {Rasmussen}\ and\ \citenamefont {Thygesen}(2015)}]{rasmussen2015computational}%
  \BibitemOpen
  \bibfield  {author} {\bibinfo {author} {\bibfnamefont {F.~A.}\ \bibnamefont {Rasmussen}}\ and\ \bibinfo {author} {\bibfnamefont {K.~S.}\ \bibnamefont {Thygesen}},\ }\bibfield  {title} {\enquote {\bibinfo {title} {Computational 2d materials database: electronic structure of transition-metal dichalcogenides and oxides},}\ }\href {https://doi.org/10.1021/acs.jpcc.5b02950} {\bibfield  {journal} {\bibinfo  {journal} {The Journal of Physical Chemistry C}\ }\textbf {\bibinfo {volume} {119}},\ \bibinfo {pages} {13169--13183} (\bibinfo {year} {2015})}\BibitemShut {NoStop}%
\bibitem [{\citenamefont {Yan}\ \emph {et~al.}(2014)\citenamefont {Yan}, \citenamefont {Qiao}, \citenamefont {Liu}, \citenamefont {Tan},\ and\ \citenamefont {Zhang}}]{yan2014photoluminescence}%
  \BibitemOpen
  \bibfield  {author} {\bibinfo {author} {\bibfnamefont {T.}~\bibnamefont {Yan}}, \bibinfo {author} {\bibfnamefont {X.}~\bibnamefont {Qiao}}, \bibinfo {author} {\bibfnamefont {X.}~\bibnamefont {Liu}}, \bibinfo {author} {\bibfnamefont {P.}~\bibnamefont {Tan}}, \ and\ \bibinfo {author} {\bibfnamefont {X.}~\bibnamefont {Zhang}},\ }\bibfield  {title} {\enquote {\bibinfo {title} {Photoluminescence properties and exciton dynamics in monolayer wse2},}\ }\href {https://doi.org/10.1063/1.4895471} {\bibfield  {journal} {\bibinfo  {journal} {Applied Physics Letters}\ }\textbf {\bibinfo {volume} {105}} (\bibinfo {year} {2014})}\BibitemShut {NoStop}%
\bibitem [{\citenamefont {He}\ \emph {et~al.}(2014)\citenamefont {He}, \citenamefont {Kumar}, \citenamefont {Zhao}, \citenamefont {Wang}, \citenamefont {Mak}, \citenamefont {Zhao},\ and\ \citenamefont {Shan}}]{he2014tightly}%
  \BibitemOpen
  \bibfield  {author} {\bibinfo {author} {\bibfnamefont {K.}~\bibnamefont {He}}, \bibinfo {author} {\bibfnamefont {N.}~\bibnamefont {Kumar}}, \bibinfo {author} {\bibfnamefont {L.}~\bibnamefont {Zhao}}, \bibinfo {author} {\bibfnamefont {Z.}~\bibnamefont {Wang}}, \bibinfo {author} {\bibfnamefont {K.~F.}\ \bibnamefont {Mak}}, \bibinfo {author} {\bibfnamefont {H.}~\bibnamefont {Zhao}}, \ and\ \bibinfo {author} {\bibfnamefont {J.}~\bibnamefont {Shan}},\ }\bibfield  {title} {\enquote {\bibinfo {title} {Tightly bound excitons in monolayer wse 2},}\ }\href {https://doi.org/10.1103/PhysRevLett.113.026803} {\bibfield  {journal} {\bibinfo  {journal} {Physical review letters}\ }\textbf {\bibinfo {volume} {113}},\ \bibinfo {pages} {026803} (\bibinfo {year} {2014})}\BibitemShut {NoStop}%
\bibitem [{\citenamefont {Vaquero}\ \emph {et~al.}(2020)\citenamefont {Vaquero}, \citenamefont {Cleric{\`o}}, \citenamefont {Salvador-S{\'a}nchez}, \citenamefont {Mart{\'\i}n-Ramos}, \citenamefont {D{\'\i}az}, \citenamefont {Dom{\'\i}nguez-Adame}, \citenamefont {Meziani}, \citenamefont {Diez},\ and\ \citenamefont {Quereda}}]{vaquero2020excitons}%
  \BibitemOpen
  \bibfield  {author} {\bibinfo {author} {\bibfnamefont {D.}~\bibnamefont {Vaquero}}, \bibinfo {author} {\bibfnamefont {V.}~\bibnamefont {Cleric{\`o}}}, \bibinfo {author} {\bibfnamefont {J.}~\bibnamefont {Salvador-S{\'a}nchez}}, \bibinfo {author} {\bibfnamefont {A.}~\bibnamefont {Mart{\'\i}n-Ramos}}, \bibinfo {author} {\bibfnamefont {E.}~\bibnamefont {D{\'\i}az}}, \bibinfo {author} {\bibfnamefont {F.}~\bibnamefont {Dom{\'\i}nguez-Adame}}, \bibinfo {author} {\bibfnamefont {Y.~M.}\ \bibnamefont {Meziani}}, \bibinfo {author} {\bibfnamefont {E.}~\bibnamefont {Diez}}, \ and\ \bibinfo {author} {\bibfnamefont {J.}~\bibnamefont {Quereda}},\ }\bibfield  {title} {\enquote {\bibinfo {title} {Excitons, trions and rydberg states in monolayer mos2 revealed by low-temperature photocurrent spectroscopy},}\ }\href {https://doi.org/10.1038/s42005-020-00460-9} {\bibfield  {journal} {\bibinfo  {journal} {Communications Physics}\ }\textbf {\bibinfo {volume} {3}},\ \bibinfo {pages} {194} (\bibinfo {year} {2020})}\BibitemShut
  {NoStop}%
\bibitem [{\citenamefont {Leisgang}\ \emph {et~al.}(2020)\citenamefont {Leisgang}, \citenamefont {Shree}, \citenamefont {Paradisanos}, \citenamefont {Sponfeldner}, \citenamefont {Robert}, \citenamefont {Lagarde}, \citenamefont {Balocchi}, \citenamefont {Watanabe}, \citenamefont {Taniguchi}, \citenamefont {Marie} \emph {et~al.}}]{leisgang2020giant}%
  \BibitemOpen
  \bibfield  {author} {\bibinfo {author} {\bibfnamefont {N.}~\bibnamefont {Leisgang}}, \bibinfo {author} {\bibfnamefont {S.}~\bibnamefont {Shree}}, \bibinfo {author} {\bibfnamefont {I.}~\bibnamefont {Paradisanos}}, \bibinfo {author} {\bibfnamefont {L.}~\bibnamefont {Sponfeldner}}, \bibinfo {author} {\bibfnamefont {C.}~\bibnamefont {Robert}}, \bibinfo {author} {\bibfnamefont {D.}~\bibnamefont {Lagarde}}, \bibinfo {author} {\bibfnamefont {A.}~\bibnamefont {Balocchi}}, \bibinfo {author} {\bibfnamefont {K.}~\bibnamefont {Watanabe}}, \bibinfo {author} {\bibfnamefont {T.}~\bibnamefont {Taniguchi}}, \bibinfo {author} {\bibfnamefont {X.}~\bibnamefont {Marie}},  \emph {et~al.},\ }\bibfield  {title} {\enquote {\bibinfo {title} {Giant stark splitting of an exciton in bilayer mos2},}\ }\href {https://doi.org/10.1038/s41565-020-0750-1} {\bibfield  {journal} {\bibinfo  {journal} {Nature nanotechnology}\ }\textbf {\bibinfo {volume} {15}},\ \bibinfo {pages} {901--907} (\bibinfo {year} {2020})}\BibitemShut {NoStop}%
\bibitem [{\citenamefont {Ugeda}\ \emph {et~al.}(2014)\citenamefont {Ugeda}, \citenamefont {Bradley}, \citenamefont {Shi}, \citenamefont {Da~Jornada}, \citenamefont {Zhang}, \citenamefont {Qiu}, \citenamefont {Ruan}, \citenamefont {Mo}, \citenamefont {Hussain}, \citenamefont {Shen} \emph {et~al.}}]{ugeda2014giant}%
  \BibitemOpen
  \bibfield  {author} {\bibinfo {author} {\bibfnamefont {M.~M.}\ \bibnamefont {Ugeda}}, \bibinfo {author} {\bibfnamefont {A.~J.}\ \bibnamefont {Bradley}}, \bibinfo {author} {\bibfnamefont {S.-F.}\ \bibnamefont {Shi}}, \bibinfo {author} {\bibfnamefont {F.~H.}\ \bibnamefont {Da~Jornada}}, \bibinfo {author} {\bibfnamefont {Y.}~\bibnamefont {Zhang}}, \bibinfo {author} {\bibfnamefont {D.~Y.}\ \bibnamefont {Qiu}}, \bibinfo {author} {\bibfnamefont {W.}~\bibnamefont {Ruan}}, \bibinfo {author} {\bibfnamefont {S.-K.}\ \bibnamefont {Mo}}, \bibinfo {author} {\bibfnamefont {Z.}~\bibnamefont {Hussain}}, \bibinfo {author} {\bibfnamefont {Z.-X.}\ \bibnamefont {Shen}},  \emph {et~al.},\ }\bibfield  {title} {\enquote {\bibinfo {title} {Giant bandgap renormalization and excitonic effects in a monolayer transition metal dichalcogenide semiconductor},}\ }\href {https://doi.org/10.1038/nmat4061} {\bibfield  {journal} {\bibinfo  {journal} {Nature materials}\ }\textbf {\bibinfo {volume} {13}},\ \bibinfo {pages} {1091--1095}
  (\bibinfo {year} {2014})}\BibitemShut {NoStop}%
\bibitem [{\citenamefont {Jasi{\'n}ski}\ \emph {et~al.}(2025)\citenamefont {Jasi{\'n}ski}, \citenamefont {Hagel}, \citenamefont {Brem}, \citenamefont {Wietek}, \citenamefont {Taniguchi}, \citenamefont {Watanabe}, \citenamefont {Chernikov}, \citenamefont {Bruyant}, \citenamefont {Dyksik}, \citenamefont {Surrente} \emph {et~al.}}]{jasinski2025quadrupolar}%
  \BibitemOpen
  \bibfield  {author} {\bibinfo {author} {\bibfnamefont {J.}~\bibnamefont {Jasi{\'n}ski}}, \bibinfo {author} {\bibfnamefont {J.}~\bibnamefont {Hagel}}, \bibinfo {author} {\bibfnamefont {S.}~\bibnamefont {Brem}}, \bibinfo {author} {\bibfnamefont {E.}~\bibnamefont {Wietek}}, \bibinfo {author} {\bibfnamefont {T.}~\bibnamefont {Taniguchi}}, \bibinfo {author} {\bibfnamefont {K.}~\bibnamefont {Watanabe}}, \bibinfo {author} {\bibfnamefont {A.}~\bibnamefont {Chernikov}}, \bibinfo {author} {\bibfnamefont {N.}~\bibnamefont {Bruyant}}, \bibinfo {author} {\bibfnamefont {M.}~\bibnamefont {Dyksik}}, \bibinfo {author} {\bibfnamefont {A.}~\bibnamefont {Surrente}},  \emph {et~al.},\ }\bibfield  {title} {\enquote {\bibinfo {title} {Quadrupolar excitons in mose2 bilayers},}\ }\href {https://doi.org/10.1038/s41467-025-56586-3} {\bibfield  {journal} {\bibinfo  {journal} {Nature Communications}\ }\textbf {\bibinfo {volume} {16}},\ \bibinfo {pages} {1382} (\bibinfo {year} {2025})}\BibitemShut {NoStop}%
\bibitem [{\citenamefont {Kyl{\"a}np{\"a}{\"a}}\ and\ \citenamefont {Komsa}(2015)}]{kylanpaa2015binding}%
  \BibitemOpen
  \bibfield  {author} {\bibinfo {author} {\bibfnamefont {I.}~\bibnamefont {Kyl{\"a}np{\"a}{\"a}}}\ and\ \bibinfo {author} {\bibfnamefont {H.-P.}\ \bibnamefont {Komsa}},\ }\bibfield  {title} {\enquote {\bibinfo {title} {Binding energies of exciton complexes in transition metal dichalcogenide monolayers and effect of dielectric environment},}\ }\href {https://doi.org/10.1103/PhysRevB.92.205418} {\bibfield  {journal} {\bibinfo  {journal} {Physical Review B}\ }\textbf {\bibinfo {volume} {92}},\ \bibinfo {pages} {205418} (\bibinfo {year} {2015})}\BibitemShut {NoStop}%
\bibitem [{\citenamefont {Wang}\ \emph {et~al.}(2019)\citenamefont {Wang}, \citenamefont {Le-Van}, \citenamefont {Vaianella}, \citenamefont {Maes}, \citenamefont {Eizagirre~Barker}, \citenamefont {Godiksen}, \citenamefont {Curto},\ and\ \citenamefont {Gomez~Rivas}}]{wang2019limits}%
  \BibitemOpen
  \bibfield  {author} {\bibinfo {author} {\bibfnamefont {S.}~\bibnamefont {Wang}}, \bibinfo {author} {\bibfnamefont {Q.}~\bibnamefont {Le-Van}}, \bibinfo {author} {\bibfnamefont {F.}~\bibnamefont {Vaianella}}, \bibinfo {author} {\bibfnamefont {B.}~\bibnamefont {Maes}}, \bibinfo {author} {\bibfnamefont {S.}~\bibnamefont {Eizagirre~Barker}}, \bibinfo {author} {\bibfnamefont {R.~H.}\ \bibnamefont {Godiksen}}, \bibinfo {author} {\bibfnamefont {A.~G.}\ \bibnamefont {Curto}}, \ and\ \bibinfo {author} {\bibfnamefont {J.}~\bibnamefont {Gomez~Rivas}},\ }\bibfield  {title} {\enquote {\bibinfo {title} {Limits to strong coupling of excitons in multilayer ws2 with collective plasmonic resonances},}\ }\href {https://doi.org/10.1021/acsphotonics.8b01459} {\bibfield  {journal} {\bibinfo  {journal} {Acs Photonics}\ }\textbf {\bibinfo {volume} {6}},\ \bibinfo {pages} {286--293} (\bibinfo {year} {2019})}\BibitemShut {NoStop}%
\bibitem [{\citenamefont {Wang}\ \emph {et~al.}(2018)\citenamefont {Wang}, \citenamefont {Chernikov}, \citenamefont {Glazov}, \citenamefont {Heinz}, \citenamefont {Marie}, \citenamefont {Amand},\ and\ \citenamefont {Urbaszek}}]{wang2018colloquium}%
  \BibitemOpen
  \bibfield  {author} {\bibinfo {author} {\bibfnamefont {G.}~\bibnamefont {Wang}}, \bibinfo {author} {\bibfnamefont {A.}~\bibnamefont {Chernikov}}, \bibinfo {author} {\bibfnamefont {M.~M.}\ \bibnamefont {Glazov}}, \bibinfo {author} {\bibfnamefont {T.~F.}\ \bibnamefont {Heinz}}, \bibinfo {author} {\bibfnamefont {X.}~\bibnamefont {Marie}}, \bibinfo {author} {\bibfnamefont {T.}~\bibnamefont {Amand}}, \ and\ \bibinfo {author} {\bibfnamefont {B.}~\bibnamefont {Urbaszek}},\ }\bibfield  {title} {\enquote {\bibinfo {title} {Colloquium: Excitons in atomically thin transition metal dichalcogenides},}\ }\href {https://doi.org/10.1103/RevModPhys.90.021001} {\bibfield  {journal} {\bibinfo  {journal} {Reviews of Modern Physics}\ }\textbf {\bibinfo {volume} {90}},\ \bibinfo {pages} {021001} (\bibinfo {year} {2018})}\BibitemShut {NoStop}%
\bibitem [{\citenamefont {Ye}\ \emph {et~al.}(2014)\citenamefont {Ye}, \citenamefont {Cao}, \citenamefont {O’brien}, \citenamefont {Zhu}, \citenamefont {Yin}, \citenamefont {Wang}, \citenamefont {Louie},\ and\ \citenamefont {Zhang}}]{ye2014probing}%
  \BibitemOpen
  \bibfield  {author} {\bibinfo {author} {\bibfnamefont {Z.}~\bibnamefont {Ye}}, \bibinfo {author} {\bibfnamefont {T.}~\bibnamefont {Cao}}, \bibinfo {author} {\bibfnamefont {K.}~\bibnamefont {O’brien}}, \bibinfo {author} {\bibfnamefont {H.}~\bibnamefont {Zhu}}, \bibinfo {author} {\bibfnamefont {X.}~\bibnamefont {Yin}}, \bibinfo {author} {\bibfnamefont {Y.}~\bibnamefont {Wang}}, \bibinfo {author} {\bibfnamefont {S.~G.}\ \bibnamefont {Louie}}, \ and\ \bibinfo {author} {\bibfnamefont {X.}~\bibnamefont {Zhang}},\ }\bibfield  {title} {\enquote {\bibinfo {title} {Probing excitonic dark states in single-layer tungsten disulphide},}\ }\href {https://doi.org/10.1038/nature13734} {\bibfield  {journal} {\bibinfo  {journal} {Nature}\ }\textbf {\bibinfo {volume} {513}},\ \bibinfo {pages} {214--218} (\bibinfo {year} {2014})}\BibitemShut {NoStop}%
\bibitem [{\citenamefont {Mai}\ \emph {et~al.}(2014)\citenamefont {Mai}, \citenamefont {Barrette}, \citenamefont {Yu}, \citenamefont {Semenov}, \citenamefont {Kim}, \citenamefont {Cao},\ and\ \citenamefont {Gundogdu}}]{mai2014many}%
  \BibitemOpen
  \bibfield  {author} {\bibinfo {author} {\bibfnamefont {C.}~\bibnamefont {Mai}}, \bibinfo {author} {\bibfnamefont {A.}~\bibnamefont {Barrette}}, \bibinfo {author} {\bibfnamefont {Y.}~\bibnamefont {Yu}}, \bibinfo {author} {\bibfnamefont {Y.~G.}\ \bibnamefont {Semenov}}, \bibinfo {author} {\bibfnamefont {K.~W.}\ \bibnamefont {Kim}}, \bibinfo {author} {\bibfnamefont {L.}~\bibnamefont {Cao}}, \ and\ \bibinfo {author} {\bibfnamefont {K.}~\bibnamefont {Gundogdu}},\ }\bibfield  {title} {\enquote {\bibinfo {title} {Many-body effects in valleytronics: direct measurement of valley lifetimes in single-layer mos2},}\ }\href {https://doi.org/10.1021/nl403742j} {\bibfield  {journal} {\bibinfo  {journal} {Nano letters}\ }\textbf {\bibinfo {volume} {14}},\ \bibinfo {pages} {202--206} (\bibinfo {year} {2014})}\BibitemShut {NoStop}%
\bibitem [{\citenamefont {Klots}\ \emph {et~al.}(2014)\citenamefont {Klots}, \citenamefont {Newaz}, \citenamefont {Wang}, \citenamefont {Prasai}, \citenamefont {Krzyzanowska}, \citenamefont {Lin}, \citenamefont {Caudel}, \citenamefont {Ghimire}, \citenamefont {Yan}, \citenamefont {Ivanov} \emph {et~al.}}]{klots2014probing}%
  \BibitemOpen
  \bibfield  {author} {\bibinfo {author} {\bibfnamefont {A.}~\bibnamefont {Klots}}, \bibinfo {author} {\bibfnamefont {A.}~\bibnamefont {Newaz}}, \bibinfo {author} {\bibfnamefont {B.}~\bibnamefont {Wang}}, \bibinfo {author} {\bibfnamefont {D.}~\bibnamefont {Prasai}}, \bibinfo {author} {\bibfnamefont {H.}~\bibnamefont {Krzyzanowska}}, \bibinfo {author} {\bibfnamefont {J.}~\bibnamefont {Lin}}, \bibinfo {author} {\bibfnamefont {D.}~\bibnamefont {Caudel}}, \bibinfo {author} {\bibfnamefont {N.}~\bibnamefont {Ghimire}}, \bibinfo {author} {\bibfnamefont {J.}~\bibnamefont {Yan}}, \bibinfo {author} {\bibfnamefont {B.}~\bibnamefont {Ivanov}},  \emph {et~al.},\ }\bibfield  {title} {\enquote {\bibinfo {title} {Probing excitonic states in suspended two-dimensional semiconductors by photocurrent spectroscopy},}\ }\href {https://doi.org/10.1021/acsphotonics.8b01459} {\bibfield  {journal} {\bibinfo  {journal} {Scientific reports}\ }\textbf {\bibinfo {volume} {4}},\ \bibinfo {pages} {6608} (\bibinfo {year} {2014})}\BibitemShut
  {NoStop}%
\bibitem [{\citenamefont {Wang}\ \emph {et~al.}(2014)\citenamefont {Wang}, \citenamefont {Bouet}, \citenamefont {Lagarde}, \citenamefont {Vidal}, \citenamefont {Balocchi}, \citenamefont {Amand}, \citenamefont {Marie},\ and\ \citenamefont {Urbaszek}}]{wang2014valley}%
  \BibitemOpen
  \bibfield  {author} {\bibinfo {author} {\bibfnamefont {G.}~\bibnamefont {Wang}}, \bibinfo {author} {\bibfnamefont {L.}~\bibnamefont {Bouet}}, \bibinfo {author} {\bibfnamefont {D.}~\bibnamefont {Lagarde}}, \bibinfo {author} {\bibfnamefont {M.}~\bibnamefont {Vidal}}, \bibinfo {author} {\bibfnamefont {A.}~\bibnamefont {Balocchi}}, \bibinfo {author} {\bibfnamefont {T.}~\bibnamefont {Amand}}, \bibinfo {author} {\bibfnamefont {X.}~\bibnamefont {Marie}}, \ and\ \bibinfo {author} {\bibfnamefont {B.}~\bibnamefont {Urbaszek}},\ }\bibfield  {title} {\enquote {\bibinfo {title} {Valley dynamics probed through charged and neutral exciton emission in monolayer wse2},}\ }\href {https://doi.org/10.1103/PhysRevB.90.075413} {\bibfield  {journal} {\bibinfo  {journal} {Physical Review B}\ }\textbf {\bibinfo {volume} {90}},\ \bibinfo {pages} {075413} (\bibinfo {year} {2014})}\BibitemShut {NoStop}%
\bibitem [{\citenamefont {Lagarde}\ \emph {et~al.}(2014)\citenamefont {Lagarde}, \citenamefont {Bouet}, \citenamefont {Marie}, \citenamefont {Zhu}, \citenamefont {Liu}, \citenamefont {Amand}, \citenamefont {Tan},\ and\ \citenamefont {Urbaszek}}]{lagarde2014carrier}%
  \BibitemOpen
  \bibfield  {author} {\bibinfo {author} {\bibfnamefont {D.}~\bibnamefont {Lagarde}}, \bibinfo {author} {\bibfnamefont {L.}~\bibnamefont {Bouet}}, \bibinfo {author} {\bibfnamefont {X.}~\bibnamefont {Marie}}, \bibinfo {author} {\bibfnamefont {C.}~\bibnamefont {Zhu}}, \bibinfo {author} {\bibfnamefont {B.}~\bibnamefont {Liu}}, \bibinfo {author} {\bibfnamefont {T.}~\bibnamefont {Amand}}, \bibinfo {author} {\bibfnamefont {P.}~\bibnamefont {Tan}}, \ and\ \bibinfo {author} {\bibfnamefont {B.}~\bibnamefont {Urbaszek}},\ }\bibfield  {title} {\enquote {\bibinfo {title} {Carrier and polarization dynamics in monolayer mos 2},}\ }\href {https://doi.org/10.1103/PhysRevLett.112.047401} {\bibfield  {journal} {\bibinfo  {journal} {Physical review letters}\ }\textbf {\bibinfo {volume} {112}},\ \bibinfo {pages} {047401} (\bibinfo {year} {2014})}\BibitemShut {NoStop}%
\bibitem [{\citenamefont {Korn}\ \emph {et~al.}(2011)\citenamefont {Korn}, \citenamefont {Heydrich}, \citenamefont {Hirmer}, \citenamefont {Schmutzler},\ and\ \citenamefont {Sch{\"u}ller}}]{korn2011low}%
  \BibitemOpen
  \bibfield  {author} {\bibinfo {author} {\bibfnamefont {T.}~\bibnamefont {Korn}}, \bibinfo {author} {\bibfnamefont {S.}~\bibnamefont {Heydrich}}, \bibinfo {author} {\bibfnamefont {M.}~\bibnamefont {Hirmer}}, \bibinfo {author} {\bibfnamefont {J.}~\bibnamefont {Schmutzler}}, \ and\ \bibinfo {author} {\bibfnamefont {C.}~\bibnamefont {Sch{\"u}ller}},\ }\bibfield  {title} {\enquote {\bibinfo {title} {Low-temperature photocarrier dynamics in monolayer mos2},}\ }\href {https://doi.org/10.1063/1.3636402} {\bibfield  {journal} {\bibinfo  {journal} {Applied Physics Letters}\ }\textbf {\bibinfo {volume} {99}} (\bibinfo {year} {2011})}\BibitemShut {NoStop}%
\end{thebibliography}%
\end{document}